\documentclass[11pt,a4paper]{article}

\usepackage[hidelinks]{hyperref}
\usepackage{amssymb, amsmath, graphicx, subcaption}
\usepackage{amsthm}
\usepackage{autobreak}
\usepackage[ruled]{algorithm2e}
\usepackage{multirow}
\usepackage{float}
\usepackage{fancyhdr}

\usepackage{geometry}
\fancypagestyle{firststyle}
{
	\fancyhf{}
	\fancyhead[RE,LO]{}
	\fancyfoot[RE,LO]{}
	\fancyfoot[LE,RO]{Page \thepage}
}

\begin{document}
	
	\title{Analytic frozen and other low eccentric orbits \\ under J2 perturbation}
	
	\author{David Arnas\thanks{Purdue University, West Lafayette, IN 47907, USA. Email: \textsc{darnas@purdue.edu}}}
	
	\date{}	
	
	\maketitle
	
	\thispagestyle{firststyle}
	
	\begin{abstract}
		This work presents an analytical perturbation method to define and study the dynamics of frozen orbits under the perturbation effects produced by the oblatness of the main celestial body. This is done using a perturbation method purely based on osculating elements. This allows to characterize, define, and study the three existing families of frozen orbits in closed-form: the two families of frozen orbits close to the critical inclination, and the family of frozen orbits that appears at low values of eccentricity. To that end, this work includes the first and second order approximate solutions of the proposed perturbation method, including their applications to define frozen orbits, repeating ground-track orbits, and sun-synchronous orbits. Examples of application are also presented to show the expected error performance of the proposed approach. 
	\end{abstract}

\section{Introduction}

One of the most important problems in astrodynamics is the study of the so called main satellite problem, that is, the analysis of the effects that the oblatness of the main celestial body, represented by the $J_2$ term of its gravitational potential, have on the evolution of an orbit. For many celestial bodies, Earth included, the $J_2$ perturbation is several orders of magnitude larger than any other perturbation affecting orbits, making this perturbation the first order approximation in many problems in astrodynamics. Unfortunately, it is well-know that the main satellite problem has no analytical solution~\cite{irigoyen1993non,celletti1995non}. Therefore, numerical and analytical approximations are required in order to study this system.

In general, numerical approximations, although simpler to implement, do not provide general results, making them less than ideal to characterize solutions. Conversely, even if they are more complex to obtain, analytical approximations provide a more powerful tool to identify, understand, and analyze different kinds of orbits. This is especially true when dealing with mission design, particularly when defining and studying frozen orbits. 

Frozen orbits are orbits that maintain several of their orbital elements in a periodic evolution, that is, they present orbital elements that repeat after a given period of time. In general, however, we refer to frozen orbits to the ones in which the semi-major axis, inclination, eccentricity, argument of perigee, and true anomaly follow the same period of repetition. This makes these orbits specially interesting for many missions, including Earth observation, telecommunications or global coverage. For this reason, this work focuses on the analytical definition and study of frozen orbits under the effects of the $J_2$ perturbation.

Early accurate approximate solutions to the main satellite problem started with the first order solutions proposed by Brower~\cite{brouwer} and Kozai~\cite{kozai1959motion}. These solutions, although very accurate with the second order solution from Kozai~\cite{kozai1962second}, presented several limitations, especially in the regions close to the critical inclination or when the eccentricity of the orbits was too small. As a result of that, Lyddane~\cite{lyddane1963small}, Cohen and Lyddane~\cite{cohen1981radius}, and Coffey, Deprit and Miller~\cite{coffey1986critical} proposed different approaches to be able to study these limit cases. Note that these specific regions are of extreme importance since they are the regions where frozen orbits appear.

With the same goal in mind, Deprit~\cite{deprit1969} introduced a new perturbation method to study the main satellite problem based on Lie series transforms. This approach allows to obtain approximate analytical solutions up to any order of precision using a canonical set of mappings of the Hamiltonian of the system. The usefulness of this approach has led to the development of the so-called Lie-Deprit methods, a set of perturbation methods that have become one of the standards to study astrodynamics problems analytically~\cite{kamel1969expansion,kamel1970perturbation,deprit1970main,deprit1982delaunay,cid,abad2001short,lara2014delaunay,mahajan2018exact,lara2019new,abad2020integration,abad2021cid}. Examples of their use include the third order solution of the main satellite problem~\cite{3rdorder} or the method of elimination of parallax~\cite{deprit1981elimination}. 

The majority of these approaches rely on different forms of averaging. This allows to significantly reduce the complexity of the resultant solutions. Additionally, since they are based on mean elements, these approaches allow to have a better understanding of the global dynamics of the problem. Nevertheless, and when studying frozen orbits, averaging techniques can present some limitations due to the singularities that can appear on their solutions in the regions close to the frozen condition. Although this does not prevent the study of these regions using these techniques, it certainly make their study more complex. This is especially true when studying bifurcation in the main satellite problem~\cite{coffey1986critical,Broucke,coffey_frozen}. 

This work instead focuses on perturbation methods based on osculating elements. Examples of these kind of methods for the main satellite problem include, for instance, the Poincar\'e-Lindstedt method~\cite{zonal}, or the use of operation theory~\cite{koopman,schur}. In here, however, we continue and expand the perturbed approximate solution proposed by Arnas~\cite{meanj2}. This has two main objectives. First, to use the solution from Ref.~\cite{meanj2} to determine and to analyze all possible frozen orbits based on an approximate second order solution. And second, to provide a modified perturbation method that simplifies the study of near circular frozen orbits, includes the time evolution of the system, and maintains the accuracy on the solution. Both objectives enable the direct study of the frozen condition from a osculating point of view, allowing the determination of the initial conditions that provide the frozen condition for the three families of frozen orbits under $J_2$. This is done using the same formalism from the perturbation method, making the analysis of these regions relatively simple once the approximate analytical solution is obtained.

This manuscript is organized as follows. First, all the frozen orbits related with an approximate second order solution under $J_2$ are identified based on the secular variation of the approximate solution from Arnas~\cite{meanj2}. Second, frozen orbits close to the critical inclination are analyzed. This includes the analytical derivation of the initial orbital elements that allow the frozen condition. Third, a modified perturbation approach is introduced to study low eccentric orbits in general, but more specifically, the family of frozen orbits that exists for values of the eccentricity in the order of magnitude of $J_2$. This approach is then used to define an analytical transformation from osculating to mean elements, and to determine the secular variation of the orbital elements, including the perturbed period of the orbit. Fourth, based on the previous results, the case of near circular frozen orbits is analyzed. Finally, three examples of application are included to show the performance of the solutions provided for each one of the different families of frozen orbits. 

All the solutions provided in this document have been translated into Matlab code for their ease of use. The resultant scripts can be found in the following web page:
\href{https://engineering.purdue.edu/ART/research/research-code}{https://engineering.purdue.edu/ART/research/research-code}.


\section{The main satellite problem}

The oblatness of a celestial body (represented by the $J_2$ component of the zonal harmonics of the gravitational potential) is usually the most important perturbation for an object orbiting that celestial body. An example of that is the Earth, where the $J_2$ perturbation is three orders of magnitude larger than any other term of the geopotential, and much larger than other perturbations but for a region very close to the Earth, where $J_2$ can be comparable with the atmospheric drag. This means that $J_2$ represents the major perturbation in many missions orbiting the Earth and determines many of the most important effects on the secular variation of orbits.

The dynamics of an orbiting object subject to $J_2$ can be analyzed through its Hamiltonian:
\begin{equation}
	\mathcal{H} = \displaystyle\frac{1}{2}\left(p_r^2 + \frac{p_{\varphi}^2}{r^2} + \frac{p_{\lambda}^2}{r^2\cos^2(\varphi)}\right) - \frac{\mu}{r} + \frac{1}{2}\mu R^2 J_2\frac{1}{r^3}\left(3\sin^2(\varphi) - 1\right),
\end{equation}
where $r$ is the radial distance from the center of the celestial body to the orbiting object, $\varphi$ is the latitude of the object with respect to the celestial body's equator, and $\lambda$ is the longitude of that object with respect to the $x$ direction in an inertial frame of reference selected where the $x$ and $y$ axis are contained in the celestial body's equatorial plane. Additionally, $\mu$ is the gravitational constant of the celestial body, $R$ is the mean equatorial radius of the celestial body, and:
\begin{eqnarray}
	p_r & = & \dot{r}; \nonumber \\
	p_{\varphi} & = & r^2\dot{\varphi}; \nonumber \\
	p_{\lambda} & = & r^2\cos^2(\varphi)\dot{\lambda};
\end{eqnarray}
are the conjugate momenta of $r$, $\varphi$ and $\lambda$ respectively. This Hamiltonian can be used to derive Hamilton's equations for the $J_2$ problem:
\begin{eqnarray}\label{eq:hamilton}
	\displaystyle\frac{dr}{dt} & = & p_r; \nonumber \\
	\displaystyle\frac{dp_r}{dt} & = & -\frac{\mu}{r^2} + \frac{p_{\varphi}^2}{r^3}  + \frac{p_{\lambda}^2}{r^3\cos^2(\varphi)} + \frac{3}{2}\mu J_2 R^2\frac{1}{r^4}\left(3\sin^2(\varphi) - 1\right); \nonumber \\
	\displaystyle\frac{d\varphi}{dt} & = & \frac{p_{\varphi}}{r^2}; \nonumber \\
	\displaystyle\frac{dp_{\varphi}}{dt} & = & -\frac{p_{\lambda}^2}{r^2}\frac{\sin(\varphi)}{\cos^3(\varphi)} - 3\mu J_2 R^2\frac{1}{r^3}\sin(\varphi)\cos(\varphi); \nonumber \\
	\displaystyle\frac{d\lambda}{dt} & = & \frac{p_{\lambda}}{r^2\cos^2(\varphi)}; \nonumber \\
	\displaystyle\frac{dp_{\lambda}}{dt} & = & 0.
\end{eqnarray}


\section{Variable transformation}

This work is based on Keplerian orbital elements: semi-major axis ($a$), eccentricity ($e$), inclination ($i$), argument of perigee ($\omega$), right ascension of the ascending node ($\Omega$), and true anomaly ($\nu$), but with some slight modifications to account for low eccentric orbits, and parabolic and hyperbolic orbits. Particularly, the orbital elements proposed by Arnas~\cite{meanj2} are used, namely: $\{A, e_x, e_y, i, \Omega, \theta\}$, where A is defined as:
\begin{equation}
    A = \left(\displaystyle\frac{R}{a(1-e^2)}\right)^2 = \left(\displaystyle\frac{\mu R \cos^2(\varphi)}{p_{\varphi}^2\cos^2(\varphi) + p_{\lambda}^2}\right)^2.
\end{equation}
Additionally, $e_x$ and $e_y$ are the two components of the eccentricity vector in the orbital plane:
\begin{eqnarray}
    e_x & = & e \cos(\omega), \nonumber \\
    e_y & = & e \sin(\omega),
\end{eqnarray}
and $\theta = \omega + \nu$ is the argument of latitude. By performing this change of variables in Eq.~\eqref{eq:hamilton} we obtain:
\begin{eqnarray} 
	\displaystyle\frac{dA}{dt} & = & 12 J_2 \sqrt[\leftroot{-1}\uproot{2}\scriptstyle 4]{\frac{\mu^2 A^{11}}{R^6}} \left(1 + e_x\cos(\theta) + e_y\sin(\theta)\right)^3 \sin(\theta) \cos(\theta) \sin^2(i); \nonumber \\
	\displaystyle\frac{de_x}{dt} & = &  \frac{3}{2} J_2 \sqrt[\leftroot{-1}\uproot{2}\scriptstyle 4]{\frac{\mu^2 A^7}{R^6}} \sin(\theta) \left(1 + e_x\cos(\theta) + e_y\sin(\theta)\right)^3 \nonumber \\
	& \times & \big(-2e_y\cos^2(i)\sin(\theta) + (1 + e_x\cos(\theta) + e_y\sin(\theta))(3\sin^2(i)\sin^2(\theta) - 1) \nonumber \\
	& - & \sin^2(i)\cos(\theta)(3e_x + 4\cos(\theta) + e_x\cos(2\theta) + e_y\sin(2\theta)))\big); \nonumber \\
	\displaystyle\frac{de_y}{dt} & = & -\frac{3}{2} J_2 \sqrt[\leftroot{-1}\uproot{2}\scriptstyle 4]{\frac{\mu^2 A^7}{R^6}} \left(1 + e_x\cos(\theta) + e_y\sin(\theta)\right)^3 \nonumber \\
	& \times & \big(2e_y\cos^3(\theta)\sin^2(i)\sin(\theta) + e_x\cos^2(\theta)(5\sin^2(i)\sin^2(\theta) - 1) \nonumber \\
	& - &  2e_x\cos^2(i)\sin^2(\theta) + \cos(\theta)(1 + e_y\sin(\theta))(7\sin^2(i)\sin^2(\theta) - 1)\big); \nonumber \\
	\displaystyle\frac{di}{dt} & = & -3 J_2 \sqrt[\leftroot{-1}\uproot{2}\scriptstyle 4]{\frac{\mu^2 A^7}{R^6}} \left(1 + e_x\cos(\theta) + e_y\sin(\theta)\right)^3 \sin(i)\cos(i)\sin(\theta)\cos(\theta); \nonumber
\end{eqnarray}
\begin{eqnarray} \label{eq:difftemp}
	\displaystyle\frac{d\Omega}{dt} & = & -3 J_2 \sqrt[\leftroot{-1}\uproot{2}\scriptstyle 4]{\frac{\mu^2 A^7}{R^6}} \left(1 + e_x\cos(\theta) + e_y\sin(\theta)\right)^3 \cos(i) \sin^2(\theta); \qquad\qquad\qquad \nonumber \\
	\displaystyle\frac{d\theta}{dt} & = &  \sqrt[\leftroot{-1}\uproot{2}\scriptstyle 4]{\frac{\mu^2 A^3}{R^6}} \left(1 + e_x\cos(\theta) + e_y\sin(\theta)\right)^2 \nonumber \\
	& \times & \left(1 + 3 J_2 A \left(1 + e_x\cos(\theta) + e_y\sin(\theta)\right)\cos^2(i) \sin^2(\theta)\right).
\end{eqnarray}

Once this is done, the time regularization proposed by Arnas~\cite{meanj2} is performed, that is, we change the independent variable from time to the argument of latitude. To be more precise, the resultant differential equation becomes:
\begin{eqnarray} \label{eq:timeregularization}
	\displaystyle\frac{dA}{d\theta} & = & 12 \frac{J_2 A^2}{\Delta} \left(1 + e_x\cos(\theta) + e_y\sin(\theta)\right) \sin(\theta) \cos(\theta) \sin^2(i); \nonumber \\
	\displaystyle\frac{de_x}{d\theta} & = &  \frac{3}{2} \frac{J_2 A}{\Delta} \sin(\theta) \left(1 + e_x\cos(\theta) + e_y\sin(\theta)\right) \nonumber \\
	& \times & \big(-2e_y\cos^2(i)\sin(\theta) + (1 + e_x\cos(\theta) + e_y\sin(\theta))(3\sin^2(i)\sin^2(\theta) - 1) \nonumber \\
	& - & \sin^2(i)\cos(\theta)(3e_x + 4\cos(\theta) + e_x\cos(2\theta) + e_y\sin(2\theta)))\big); \nonumber \\
	\displaystyle\frac{de_y}{d\theta} & = & -\frac{3}{2} \frac{J_2 A}{\Delta} \left(1 + e_x\cos(\theta) + e_y\sin(\theta)\right) \nonumber \\
	& \times & \big(2e_y\cos^3(\theta)\sin^2(i)\sin(\theta) + e_x\cos^2(\theta)(5\sin^2(i)\sin^2(\theta) - 1) \nonumber \\
	& - &  2e_x\cos^2(i)\sin^2(\theta) + \cos(\theta)(1 + e_y\sin(\theta))(7\sin^2(i)\sin^2(\theta) - 1)\big); \nonumber \\
	\displaystyle\frac{di}{d\theta} & = & -3 \frac{J_2 A}{\Delta} \left(1 + e_x\cos(\theta) + e_y\sin(\theta)\right) \sin(i)\cos(i)\sin(\theta)\cos(\theta); \nonumber \\
	\displaystyle\frac{d\Omega}{d\theta} & = & -3 \frac{J_2 A}{\Delta} \left(1 + e_x\cos(\theta) + e_y\sin(\theta)\right) \cos(i) \sin^2(\theta); \nonumber \\
	\displaystyle\frac{dt}{d\theta} & = &  \sqrt[\leftroot{-1}\uproot{2}\scriptstyle 4]{\frac{R^6}{\mu^2 A^3}}\frac{1}{\Delta \left(1 + e_x\cos(\theta) + e_y\sin(\theta)\right)^2}.
\end{eqnarray}
where:
\begin{equation}
    \Delta = 1 + 3 J_2 A \left(1 + e_x\cos(\theta) + e_y\sin(\theta)\right)\cos^2(i) \sin^2(\theta).
\end{equation}
This is the set of equations that is used in this work to both find analytical approximations to the system, and to find frozen orbits under $J_2$.


\section{Finding frozen orbits from the second order \\ osculating solution}
\label{sec:generalfrozen}

In this work we consider the frozen condition as the one that allows orbits to have a periodic evolution of their variables $\{A, e_x, e_y, i\}$ after one complete orbital revolution, that is, since the independent variable in Eq.~\eqref{eq:timeregularization} is the argument of latitude, this is equivalent to:
\begin{eqnarray}
    A(\theta = \theta_0) & = & A(\theta = \theta_0 + 2\pi); \nonumber \\
    e_x(\theta = \theta_0) & = & e_x(\theta = \theta_0 + 2\pi); \nonumber \\
    e_y(\theta = \theta_0) & = & e_y(\theta = \theta_0 + 2\pi); \nonumber \\
    i(\theta = \theta_0) & = & i(\theta = \theta_0 + 2\pi),
\end{eqnarray}
where $\theta_0$ is the initial argument of latitude of the orbit. Therefore, the first objective is to characterize the possible frozen orbits under $J_2$. This is done using the results from the perturbation approach shown in Arnas~\cite{meanj2}. Particularly, the secular variation of variables $A$ and $i$ using up to a second order expansion is:
\begin{eqnarray}
    \Delta\left.A\right|_{sec} & = & -\frac{3}{4} J_2^2 A_0^3 \pi \sin^2(i_0) (13 e_{x0} e_{y0} - 12 (e_{y0} \cos^3(\theta_0) + e_{x0} \sin^3(\theta_0)) \nonumber \\
    & + & 5 \cos(2 i_0) (-4 e_{y0} \cos^3(\theta_0) + e_{x0} (3 e_{y0} - 3 \sin(\theta_0) + \sin(3 \theta_0)))); \nonumber \\
    \Delta\left.i\right|_{sec} & = & \frac{3}{64} J_2^2 A_0^2 \pi (5 \sin(4 i_0) (-4 e_{y0} \cos^3(\theta_0) + e_{x0} (3 e_{y0} - 3 \sin(\theta_0) + \sin(3 \theta_0))) \nonumber \\
    & + & 2 \sin(2 i_0) (-12 e_{y0} \cos^3(\theta_0) + e_{x0} (13 e_{y0} - 9 \sin(\theta_0) + 3 \sin(3 \theta_0)))),
\end{eqnarray}
where $\{A_0, e_{x0}, e_{y0}, i_0, \theta_0\}$ are the initial orbital elements of the orbit. On the other hand, the secular variation of the components of the eccentricity vector are:
\begin{eqnarray}
   \Delta\left.e_x\right|_{sec} & = & -\frac{3}{4} J_2 A_0 \pi e_{y0} (3 + 5 \cos(2 i_0)) - \frac{3}{1024} J_2^2 A_0^2 \pi  (280 \cos (2 i_0) e_{y0}^3\nonumber \\
    & - & 630 \cos (4 i_0) e_{y0}^3+350 e_{y0}^3-36
   \sin (2 i_0-5 \theta_0) e_{y0}^2+45 \sin (4 i_0-5 \theta_0) e_{y0}^2\nonumber \\
    & - & 380 \sin (2
   i_0-3 \theta_0) e_{y0}^2+55 \sin (4 i_0-3 \theta_0) e_{y0}^2+1320 \sin (2
   i_0-\theta_0) e_{y0}^2\nonumber \\
    & - & 390 \sin (4 i_0-\theta_0) e_{y0}^2-1980 \sin (\theta_0)
   e_{y0}^2+630 \sin (3 \theta_0) e_{y0}^2+18 \sin (5 \theta_0) e_{y0}^2\nonumber \\
    & - & 1320 \sin (2
   i_0+\theta_0) e_{y0}^2+390 \sin (4 i_0+\theta_0) e_{y0}^2+380 \sin (2 i_0+3
   \theta_0) e_{y0}^2\nonumber \\
    & - & 55 \sin (4 i_0+3 \theta_0) e_{y0}^2+36 \sin (2 i_0+5 \theta_0)
   e_{y0}^2-45 \sin (4 i_0+5 \theta_0) e_{y0}^2\nonumber \\
    & + & 150 e_{x0}^2 e_{y0}+952 e_{x0}^2 \cos (2 i_0) e_{y0}+592
   \cos (2 i_0) e_{y0}+690 e_{x0}^2 \cos (4 i_0) e_{y0}\nonumber \\
    & - & 820 \cos (4 i_0) e_{y0}+72 e_{x0} \cos (2 i_0-5
   \theta_0) e_{y0}-90 e_{x0} \cos (4 i_0-5 \theta_0) e_{y0}\nonumber \\
    & + & 616 e_{x0} \cos (2 i_0-3 \theta_0)
   e_{y0}-50 e_{x0} \cos (4 i_0-3 \theta_0) e_{y0}\nonumber \\
    & + & 144 \cos (2 (i_0-2 \theta_0)) e_{y0}-360 \cos
   (4 i_0-2 \theta_0) e_{y0}\nonumber \\
    & + & 1056 \cos (2 (i_0-\theta_0)) e_{y0}-180 \cos (4
   (i_0-\theta_0)) e_{y0}\nonumber \\
    & - & 816 e_{x0} \cos (2 i_0-\theta_0) e_{y0}-1620 e_{x0} \cos (4
   i_0-\theta_0) e_{y0}-1272 e_{x0} \cos (\theta_0) e_{y0}\nonumber \\
    & + & 1680 \cos (2 \theta_0) e_{y0}+916
   e_{x0} \cos (3 \theta_0) e_{y0}+72 \cos (4 \theta_0) e_{y0}\nonumber \\
    & + & 36 e_{x0} \cos (5 \theta_0)
   e_{y0}+1056 \cos (2 (i_0+\theta_0)) e_{y0}-180 \cos (4 (i_0+\theta_0)) e_{y0}\nonumber \\
    & - & 816 e_{x0} \cos
   (2 i_0+\theta_0) e_{y0}-360 \cos (2 (2 i_0+\theta_0)) e_{y0}\nonumber \\
    & - & 1620 e_{x0} \cos (4
   i_0+\theta_0) e_{y0}+144 \cos (2 (i_0+2 \theta_0)) e_{y0}\nonumber \\
    & + & 616 e_{x0} \cos (2 i_0+3
   \theta_0) e_{y0}-50 e_{x0} \cos (4 i_0+3 \theta_0) e_{y0}\nonumber \\
    & + & 72 e_{x0} \cos (2 i_0+5 \theta_0)
   e_{y0}-90 e_{x0} \cos (4 i_0+5 \theta_0) e_{y0}+228 e_{y0}+2064 \pi  e_{x0}\nonumber \\
    & + & 2880 \pi  e_{x0} \cos (2
   i_0)+1200 \pi  e_{x0} \cos (4 i_0)+36 e_{x0}^2 \sin (2 i_0-5 \theta_0)\nonumber \\
    & - & 45 e_{x0}^2 \sin (4
   i_0-5 \theta_0)+236 e_{x0}^2 \sin (2 i_0-3 \theta_0)+112 \sin (2 i_0-3
   \theta_0)\nonumber \\
    & + & 5 e_{x0}^2 \sin (4 i_0-3 \theta_0)-140 \sin (4 i_0-3 \theta_0)+144
   e_{x0} \sin (2 (i_0-2 \theta_0))\nonumber \\
    & + & 360 e_{x0} \sin (4 i_0-2 \theta_0)+672 e_{x0} \sin (2
   (i_0-\theta_0))-180 e_{x0} \sin (4 (i_0-\theta_0))\nonumber \\
    & + & 72 e_{x0}^2 \sin (2
   i_0-\theta_0)+624 \sin (2 i_0-\theta_0)+210 e_{x0}^2 \sin (4 i_0-\theta_0)\nonumber \\
    & + & 420
   \sin (4 i_0-\theta_0)-204 e_{x0}^2 \sin (\theta_0)-984 \sin (\theta_0)-1008
   e_{x0} \sin (2 \theta_0)\nonumber \\
    & - & 286 e_{x0}^2 \sin (3 \theta_0)-56 \sin (3
   \theta_0)-72 e_{x0} \sin (4 \theta_0)-18 e_{x0}^2 \sin (5 \theta_0)\nonumber \\ 
   & - & 672 e_{x0} \sin
   (2 (i_0+\theta_0))+180 e_{x0} \sin (4 (i_0+\theta_0))-72 e_{x0}^2 \sin (2
   i_0+\theta_0)\nonumber \\
    & - & 624 \sin (2 i_0+\theta_0)-360 e_{x0} \sin (2 (2
   i_0+\theta_0))-210 e_{x0}^2 \sin (4 i_0+\theta_0)\nonumber
\end{eqnarray}
\begin{eqnarray}
    & - & 420 \sin (4 i_0+\theta_0)-144
   e_{x0} \sin (2 (i_0+2 \theta_0))-236 e_{x0}^2 \sin (2 i_0+3 \theta_0)\nonumber \\
    & - & 112 \sin (2
   i_0+3 \theta_0)-5 e_{x0}^2 \sin (4 i_0+3 \theta_0)+140 \sin (4 i_0+3
   \theta_0)\nonumber \\
    & - & 36 e_{x0}^2 \sin (2 i_0+5 \theta_0)+45 e_{x0}^2 \sin (4 i_0+5
   \theta_0)),
\end{eqnarray}
and:
\begin{eqnarray}
    \Delta\left.e_y\right|_{sec} & = & \displaystyle\frac{3}{4} J_2 \pi A_0 e_{x0} (3 + 5 \cos(2 i_0)) -\displaystyle\frac{3}{1024} J_2^2 \pi A_0^2 (92 e_{x0} - 326 e_{x0}^3 \nonumber \\
     & + & 2064 \pi e_{y0} - 526 e_{x0} e_{y0}^2 - 
    8 (127 e_{x0}^3 - 360 \pi e_{y0} \nonumber \\
     & + & e_{x0} (-6 + 43 e_{y0}^2)) \cos(2 i_0) - 
    10 (14 e_{x0} + 45 e_{x0}^3 - 120 \pi e_{y0} \nonumber \\
     & - & 87 e_{x0} e_{y0}^2) \cos(4 i_0) + 
    36 e_{x0}^2 \cos(2 i_0 - 5 \theta_0) - 36 e_{y0}^2 \cos(2 i_0 - 5 \theta_0) \nonumber \\
     & - & 
    45 e_{x0}^2 \cos(4 i_0 - 5 \theta_0) + 45 e_{y0}^2 \cos(4 i_0 - 5 \theta_0) + 
    112 \cos(2 i_0 - 3 \theta_0) \nonumber \\
     & - & 44 e_{x0}^2 \cos(2 i_0 - 3 \theta_0) - 
    100 e_{y0}^2 \cos(2 i_0 - 3 \theta_0) - 140 \cos(4 i_0 - 3 \theta_0) \nonumber \\
     & + & 
    235 e_{x0}^2 \cos(4 i_0 - 3 \theta_0) - 175 e_{y0}^2 \cos(4 i_0 - 3 \theta_0) \nonumber \\
     & + &
    144 e_{x0} \cos(2 (i_0 - 2 \theta_0)) + 1080 e_{x0} \cos(4 i_0 - 2 \theta_0) \nonumber \\
     & + & 
    288 e_{x0} \cos(2 (i_0 - \theta_0)) - 180 e_{x0} \cos(4 (i_0 - \theta_0)) + 
    720 \cos(2 i_0 - \theta_0) \nonumber \\
     & + & 1224 e_{x0}^2 \cos(2 i_0 - \theta_0) + 
    264 e_{y0}^2 \cos(2 i_0 - \theta_0) + 300 \cos(4 i_0 - \theta_0) \nonumber \\
     & + & 
    1410 e_{x0}^2 \cos(4 i_0 - \theta_0) - 30 e_{y0}^2 \cos(4 i_0 - \theta_0) + 1032 \cos(\theta_0) \nonumber \\
     & + & 
    1644 e_{x0}^2 \cos(\theta_0) + 300 e_{y0}^2 \cos(\theta_0) + 336 e_{x0} \cos(2 \theta_0) + 
    56 \cos(3 \theta_0) \nonumber \\
     & - & 126 e_{x0}^2 \cos(3 \theta_0) - 218 e_{y0}^2 \cos(3 \theta_0) + 
    72 e_{x0} \cos(4 \theta_0) + 18 e_{x0}^2 \cos(5 \theta_0) \nonumber \\
     & - & 18 e_{y0}^2 \cos(5 \theta_0) + 
    288 e_{x0} \cos(2 (i_0 + \theta_0)) - 180 e_{x0} \cos(4 (i_0 + \theta_0)) \nonumber \\
     & + & 
    720 \cos(2 i_0 + \theta_0) + 1224 e_{x0}^2 \cos(2 i_0 + \theta_0) + 
    264 e_{y0}^2 \cos(2 i_0 + \theta_0) \nonumber \\
     & + & 1080 e_{x0} \cos(2 (2 i_0 + \theta_0)) + 
    300 \cos(4 i_0 + \theta_0) + 1410 e_{x0}^2 \cos(4 i_0 + \theta_0) \nonumber \\
     & - & 
    30 e_{y0}^2 \cos(4 i_0 + \theta_0) + 144 e_{x0} \cos(2 (i_0 + 2 \theta_0)) + 
    112 \cos(2 i_0 + 3 \theta_0) \nonumber \\
     & - & 44 e_{x0}^2 \cos(2 i_0 + 3 \theta_0) - 
    100 e_{y0}^2 \cos(2 i_0 + 3 \theta_0) - 140 \cos(4 i_0 + 3 \theta_0) \nonumber \\
     & + & 
    235 e_{x0}^2 \cos(4 i_0 + 3 \theta_0) - 175 e_{y0}^2 \cos(4 i_0 + 3 \theta_0) + 
    36 e_{x0}^2 \cos(2 i_0 + 5 \theta_0) \nonumber \\
     & - & 36 e_{y0}^2 \cos(2 i_0 + 5 \theta_0) - 
    45 e_{x0}^2 \cos(4 i_0 + 5 \theta_0) + 45 e_{y0}^2 \cos(4 i_0 + 5 \theta_0) \nonumber \\
     & - & 
    72 e_{x0} e_{y0} \sin(2 i_0 - 5 \theta_0) + 90 e_{x0} e_{y0} \sin(4 i_0 - 5 \theta_0) \nonumber \\
     & - & 
    56 e_{x0} e_{y0} \sin(2 i_0 - 3 \theta_0) - 410 e_{x0} e_{y0} \sin(4 i_0 - 3 \theta_0) \nonumber \\
     & - & 
    144 e_{y0} \sin(2 (i_0 - 2 \theta_0)) - 360 e_{y0} \sin(4 i_0 - 2 \theta_0) \nonumber \\
     & - & 
    672 e_{y0} \sin(2 (i_0 - \theta_0)) + 180 e_{y0} \sin(4 (i_0 - \theta_0)) \nonumber \\
     & - & 
    1392 e_{x0} e_{y0} \sin(2 i_0 - \theta_0) + 780 e_{x0} e_{y0} \sin(4 i_0 - \theta_0) + 
    1848 e_{x0} e_{y0} \sin(\theta_0) \nonumber \\
     & + & 1008 e_{y0} \sin(2 \theta_0) + 92 e_{x0} e_{y0} \sin(3 \theta_0) + 
    72 e_{y0} \sin(4 \theta_0) \nonumber \\
     & + & 36 e_{x0} e_{y0} \sin(5 \theta_0) + 672 e_{y0} \sin(2 (i_0 + \theta_0)) - 
    180 e_{y0} \sin(4 (i_0 + \theta_0)) \nonumber \\
     & + & 1392 e_{x0} e_{y0} \sin(2 i_0 + \theta_0) + 
    360 e_{y0} \sin(2 (2 i_0 + \theta_0)) \nonumber \\
     & - & 780 e_{x0} e_{y0} \sin(4 i_0 + \theta_0) + 
    144 e_{y0} \sin(2 (i_0 + 2 \theta_0)) \nonumber \\
     & + & 56 e_{x0} e_{y0} \sin(2 i_0 + 3 \theta_0) + 
    410 e_{x0} e_{y0} \sin(4 i_0 + 3 \theta_0) \nonumber \\
     & + & 72 e_{x0} e_{y0} \sin(2 i_0 + 5 \theta_0) - 
    90 e_{x0} e_{y0} \sin(4 i_0 + 5 \theta_0)).
\end{eqnarray}

The frozen condition can be obtained using the previous set of equations. Particularly, and when analyzing the first order terms of the secular variation in the components of the eccentricity vector, it is easy to observe that the only possibilities to banish these terms are: 
\begin{itemize}
    \item Low eccentric orbits for any inclination but for a region close to the critical inclination. This corresponds to the condition $\mathcal{O}(e_0) = \mathcal{O}(J_2)$.
    \item Orbits with inclination close to the critical inclination. This corresponds to the condition $\mathcal{O}(3 + 5 \cos(2 i_0)) = \mathcal{O}(J_2)$.
\end{itemize}
Low eccentric orbits are discussed in more detail starting in Section~\ref{sec:perturbation}, where a modified perturbation approach is used to account for the particularities of these orbits and to reduce the length of the resultant expressions. On the other hand, orbits close to the critical inclination are studied in Section~\ref{sec:critical}.

\section{Frozen orbits close to the critical inclination} \label{sec:critical}

The first order secular terms in $A$, $i$, $e_x$, and $e_y$ banish if $\mathcal{O}(3 + 5 \cos(2 i_0)) = \mathcal{O}(J_2)$, being the only remaining secular effects the ones associated with the second order solution. However, analyzing the complete expression directly can lead to several singularities due to the close presence of the critical inclination. Nevertheless, it is still possible to analyze this region by introducing a change of variable such that:
\begin{equation} \label{eq:critexp}
    K J_2 = 3 + 5 \cos(2 i_0).
\end{equation}
This change of variable can be applied to the secular variations of $A$, $i$, $e_x$, and $e_y$ to study the second order terms of these equations. Particularly, by using this expansion, it is possible to obtain the following approximation:
\begin{eqnarray}
    \Delta\left.A\right|_{sec} & = & -3 J_2^2 A_0^3 \pi e_{x0} e_{y0} \sin^2(i_0)  + \mathcal{O}(J_2^3); \nonumber \\
    \Delta\left.i\right|_{sec} & = & \frac{3}{8} J_2^2 A_0^2 \pi e_{x0} e_{y0} \sin(2 i_0)  + \mathcal{O}(J_2^3); \nonumber \\
    \Delta\left.e_x\right|_{sec} & = & -\frac{3}{20} J_2^2 A_0 \pi e_{y0} (2 A_0 + 5 K - 12 A_0 e_{x0}^2 + 7 A_0 e_{y0}^2 + 12 A_0 e_{x0} \cos(\theta_0) \nonumber \\
    & + & 12 A_0 \cos(2 \theta_0) + 4 A_0 e_{x0} \cos(3 \theta_0) -  12 A_0 e_{y0} \sin(\theta_0) + 
   4 A_0 e_{y0} \sin(3 \theta_0)) \nonumber \\
    & + & \mathcal{O}(J_2^3); \nonumber \\
    \Delta\left.e_y\right|_{sec} & = & -\frac{3}{20} J_2^2 A_0 \pi e_{x0} (2 A_0 - 5 K + 8 A_0 e_{x0}^2 - 11 A_0 e_{y0}^2 - 12 A_0 e_{x0} \cos(\theta_0) \nonumber \\
    & - & 12 A_0 \cos(2 \theta_0) - 4 A_0 e_{x0} \cos(3 \theta_0) + 12 A_0 e_{y0} \sin(\theta_0) - 
   4 A_0 e_{y0}  \sin(3 \theta_0)) \nonumber \\
    & + & \mathcal{O}(J_2^3).
\end{eqnarray}
This means that in order to banish all the second order secular effects in $A$ and $i$, the product $\mathcal{O}(e_{x0} e_{y0}) = \mathcal{O}(J_2)$, that is, either $\mathcal{O}(e_{x0}) = \mathcal{O}(J_2)$, or $\mathcal{O}(e_{y0}) = \mathcal{O}(J_2)$, or both conditions at the same time. Since the case $\mathcal{O}(e_{x0}) = \mathcal{O}(e_{y0}) = \mathcal{O}(J_2)$ is covered in the next sections, we focus on the other two possibilities in this section.

\subsection{Case $\mathcal{O}(e_{x0}) = \mathcal{O}(J_2)$}

If $\mathcal{O}(e_{x0}) = \mathcal{O}(J_2)$, the second order secular effects on the orbital elements are:
\begin{eqnarray}
    \Delta\left.A\right|_{sec} & = & 0 + \mathcal{O}(J_2^3); \nonumber \\
    \Delta\left.i\right|_{sec} & = & 0 + \mathcal{O}(J_2^3); \nonumber \\
    \Delta\left.e_x\right|_{sec} & = & -\frac{3}{20} J_2^2 A_0 \pi e_{y0} (2 A_0 + 5 K + 7 A_0 e_{y0}^2 + 12 A_0 \cos(2 \theta_0)  -  12 A_0 e_{y0} \sin(\theta_0) \nonumber \\
    & + & 4 A_0 e_{y0} \sin(3 \theta_0)) + \mathcal{O}(J_2^3); \nonumber \\
    \Delta\left.e_y\right|_{sec} & = & 0 + \mathcal{O}(J_2^3).
\end{eqnarray}
Therefore, the condition for frozen orbit happens when:
\begin{equation}
    2 A_0 + 5 K + 7 A_0 e_{y0}^2 + 12 A_0 \cos(2 \theta_0)  -  12 A_0 e_{y0} \sin(\theta_0) + 4 A_0 e_{y0} \sin(3 \theta_0) = 0,
\end{equation}
that is, for each combination of initial $A_0$, and $i_0$ there are two potential values of the $y$ component of the eccentricity vector $y_0$ that make the orbit frozen. In addition, it can be noted that there is no specific value for the $x$ component of the eccentricity vector $x_0$, apart from the fact that $\mathcal{O}(e_{x0}) = \mathcal{O}(J_2)$. This means that the frozen condition can be achieved, at second order, as long as the orbit has an argument of perigee close to either $\pi/2$ or $3\pi/2$ radians, and with an initial value of $e_{y0}$ equal to:
\begin{eqnarray}
    e_{y0} & = & \frac{6}{7}\sin (\theta_0) - \frac{2}{7}\sin (3 \theta_0) \nonumber \\
     & \pm &  \frac{1}{7}\sqrt{6 - 35 \frac{K}{A_0} - 114 \cos (2 \theta_0) + 12 \cos (4 \theta_0) - 2 \cos(6 \theta_0)}.
\end{eqnarray}
Alternatively, the inclination of the orbit can be obtained as a function of the initial eccentricity:
\begin{equation}
    K = - \frac{1}{5} A_0 (2 + 7 e_{y0}^2 + 12 \cos(2 \theta_0) - 12 e_{y0} \sin(\theta_0) + 
   4 e_{y0} \sin(3 \theta_0)),
\end{equation}
which corresponds with the initial inclination:
\begin{equation}
    i_0 = \displaystyle\frac{1}{2}\arccos{\Big(-\displaystyle\frac{1}{25} J_2 A_0 (2 + 7 e_{y0}^2 + 12 \cos(2 \theta_0) - 12 e_{y0} \sin(\theta_0) + 
   4 e_{y0} \sin(3 \theta_0)) - \frac{3}{5}\Big)},
\end{equation}
or in terms of the initial semi-major axis of the orbit:
\begin{equation}
    i_0 = \displaystyle\frac{1}{2}\arccos{\Big(-\displaystyle\frac{1}{25} J_2 R^2 \frac{(2 + 7 e_{y0}^2 + 12 \cos(2 \theta_0) - 12 e_{y0} \sin(\theta_0) + 
   4 e_{y0} \sin(3 \theta_0))}{a_0^2(1 - e_{y0}^2)^2} - \frac{3}{5}\Big)}.
\end{equation}

\subsection{Case $\mathcal{O}(e_{y0}) = \mathcal{O}(J_2)$}

If instead, $\mathcal{O}(e_{y0}) = \mathcal{O}(J_2)$, the second order secular effects on the orbital elements are:
\begin{eqnarray}
    \Delta\left.A\right|_{sec} & = & 0 + \mathcal{O}(J_2^3); \nonumber \\
    \Delta\left.i\right|_{sec} & = & 0 + \mathcal{O}(J_2^3); \nonumber \\
    \Delta\left.e_x\right|_{sec} & = &  0 + \mathcal{O}(J_2^3); \nonumber \\
    \Delta\left.e_y\right|_{sec} & = & -\frac{3}{20} J_2^2 A_0 \pi e_{x0} (2 A_0 - 5 K + 8 A_0 e_{x0}^2 - 12 A_0 \cos(2 \theta_0)  -  12 A_0 e_{x0} \cos(\theta_0) \nonumber \\
    & - & 4 A_0 e_{y0} \cos(3 \theta_0)) + \mathcal{O}(J_2^3);.
\end{eqnarray}
This implies that the condition for frozen orbit happens when the $y$ component of the eccentricity vector is of an order of magnitude comparable with $J_2$ but without an specific value as seen in the previous case. In addition, the initial value of the $x$ component of the eccentricity vector must fulfill the following condition:
\begin{equation}
    2 A_0 - 5 K + 8 A_0 e_{x0}^2 - 12 A_0 \cos(2 \theta_0)  -  12 A_0 e_{x0} \cos(\theta_0) - 4 A_0 e_{y0} \cos(3 \theta_0) = 0,
\end{equation}
whose solutions are:
\begin{eqnarray}
    e_{x0} & = & \frac{3}{4}\cos (\theta_0) - \frac{1}{4}\cos (3 \theta_0) \nonumber \\
     & \pm &  \frac{\sqrt{2}}{8}\sqrt{2 + 20 \frac{K}{A_0} + 63 \cos (2 \theta_0) + 6 \cos (4 \theta_0) + \cos(6 \theta_0)}.
\end{eqnarray}
Additionally, we can express this frozen condition applied to the inclination of the orbit. Particularly: 
\begin{equation}
    K = - \frac{2}{5} A_0 (-1 - 4 e_{x0}^2 + 6 e_{x0} \cos(\theta_0) + 6 \cos(2\theta_0) + 2 e_{x0} \cos(3 \theta_0)),
\end{equation}
which corresponds with the initial inclination:
\begin{equation}
    i_0 = \displaystyle\frac{1}{2}\arccos{\left(-\displaystyle\frac{2}{25} J_2 A_0 (-1 - 4 e_{x0}^2 + 6 e_{x0} \cos(\theta_0) + 6 \cos(2\theta_0) + 2 e_{x0} \cos(3 \theta_0)) - \frac{3}{5}\right)},
\end{equation}
or in terms of the initial semi-major axis of the orbit:
\begin{equation}
    i_0 = \displaystyle\frac{1}{2}\arccos{\left(-\displaystyle\frac{2}{25} J_2 R^2 \frac{(-1 - 4 e_{x0}^2 + 6 e_{x0} \cos(\theta_0) + 6 \cos(2\theta_0) + 2 e_{x0} \cos(3 \theta_0))}{a_0^2(1 - e_{y0}^2)^2} - \frac{3}{5}\right)}.
\end{equation}

\begin{figure}[h!]
	\centering
	{\includegraphics[width = 0.8\textwidth]{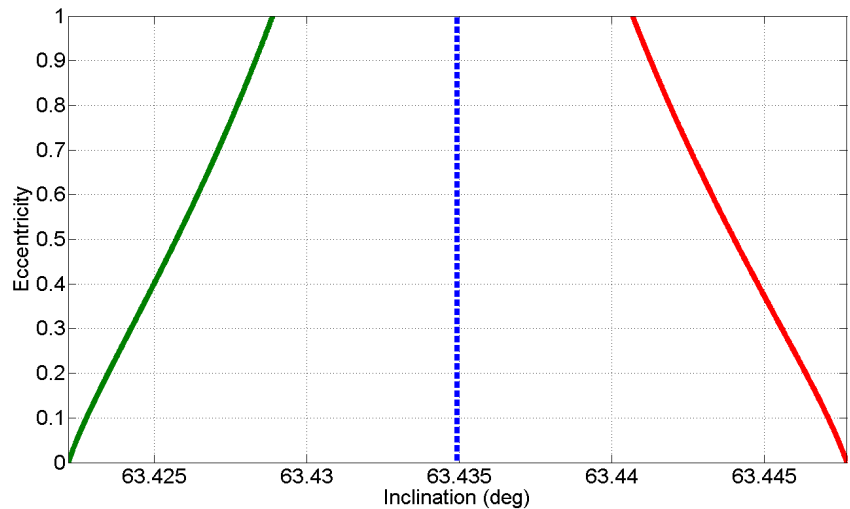}}
	\caption{Osculating value map in eccentricity and inclination for frozen orbits close to the critical inclination.}
	\label{fig:bif_osc}
\end{figure}

\begin{figure}[h!]
	\centering
	{\includegraphics[width = 0.8\textwidth]{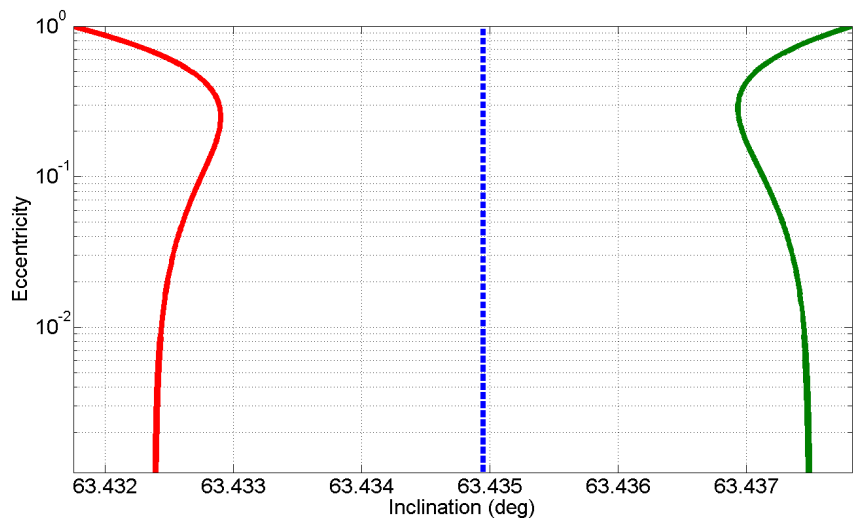}}
	\caption{Mean value map in eccentricity and inclination for frozen orbits close to the critical inclination.}
	\label{fig:bif_mean}
\end{figure}

As an example of application, Fig.~\ref{fig:bif_osc} shows the osculating evolution of the eccentricity and inclination for frozen orbits when the value of the orbital periapsis is fixed at 650 km of altitude over the Earth's surface. The curve on the right represents the solution for the case $\mathcal{O}(e_{y0}) = \mathcal{O}(J_2)$ with $\theta_0 = 0$ deg, while the curve on the left represent the frozen orbits for $\mathcal{O}(e_{x0}) = \mathcal{O}(J_2)$ with $\theta_0 = 90$ deg. The dashed line is the position of the critical inclination. As can be seen, the osculating inclination is close to the critical inclination, however, it is not the critical inclination for all the orbits included, not even in mean values. To show this better, Fig.~\ref{fig:bif_mean} shows the mean value of the eccentricity and inclination for these orbits using the analytical transformation proposed in Ref.~\cite{meanj2}, where the left-most curve is related with the case $\mathcal{O}(e_{y0}) = \mathcal{O}(J_2)$, and the right one with the case $\mathcal{O}(e_{x0}) = \mathcal{O}(J_2)$. Note that this figure clearly shows that no single frozen orbit has a mean inclination equal to the critical inclination. Additionally, it is possible to observe bifurcation appearing from the critical inclination, as predicted and analyzed by many other authors using different approaches~\cite{coffey1986critical,Broucke,coffey_frozen}. In this regard, it is important to remember that these frozen orbits have been defined by pure analytical methods and it do not require any additional analysis to study the bifurcation. This shows a very good application for the perturbation method proposed.

\section{Perturbation method for low eccentric orbits} \label{sec:perturbation}

As it was shown in the previous section, in the $J_2$ problem there is an interesting family of frozen orbits that appears at eccentricities with an order of magnitude comparable with $J_2$. Therefore, in here we use this result to provide a perturbation method to account for this property. This allows to simplify the resultant analytical expressions from Ref.~\cite{meanj2}, and provides a deeper insight into the problem. Particularly, this allows to also study the time evolution of the system and to obtain the frozen condition for any inclination.

In the case of frozen orbits and, in general, orbits whose eccentricity value is small (an order of magnitude similar to $J_2$), it is possible to introduce this information into the perturbation method to simplify the resultant transformations and remove all potential singularities when finding frozen orbits. Particularly, in these kind of orbits, both eccentricity components, $e_{x}$ and $e_{y}$, are on the order of magnitude of $J_2$, and thus, can also be expanded in the differential equation in powers of the small parameter in the same way it was done for the case of $J_2$ in Arnas~\cite{meanj2}. In order to do that, we define the following change of variables:
\begin{eqnarray}
    X = \displaystyle\frac{e_x}{J_2}, \nonumber \\
    Y = \displaystyle\frac{e_y}{J_2},
\end{eqnarray}
and introduce it in the differential equation to obtain:
\begin{eqnarray}
	\displaystyle\frac{dA}{d\theta} & = & 12 \frac{J_2 A^2}{\Delta} \left(1 + J_2 X\cos(\theta) + J_2 Y\sin(\theta)\right) \sin(\theta) \cos(\theta) \sin^2(i); \nonumber \\
	\displaystyle\frac{dX}{d\theta} & = &  \frac{3}{2} \frac{A}{\Delta} \sin(\theta) \left(1 + J_2 X\cos(\theta) + J_2 Y\sin(\theta)\right) \big(-2 J_2 Y\cos^2(i)\sin(\theta) \nonumber \\
	& + &  (1 + J_2 X\cos(\theta) + J_2 Y\sin(\theta))(3\sin^2(i)\sin^2(\theta) - 1) \nonumber \\
	& - & \sin^2(i)\cos(\theta)(3 J_2 X + 4\cos(\theta) + J_2 X \cos(2\theta) + J_2 Y \sin(2\theta)))\big); \nonumber \\
	\displaystyle\frac{dY}{d\theta} & = & -\frac{3}{2} \frac{A}{\Delta} \left(1 + J_2 X\cos(\theta) + J_2 Y\sin(\theta)\right) \nonumber \\
	& \times & \big(2 J_2 Y\cos^3(\theta)\sin^2(i)\sin(\theta) + J_2 X \cos^2(\theta)(5\sin^2(i)\sin^2(\theta) - 1) \nonumber \\
	& - &  2 J_2 X \cos^2(i)\sin^2(\theta) + \cos(\theta)(1 + J_2 Y \sin(\theta))(7\sin^2(i)\sin^2(\theta) - 1)\big); \nonumber \\
	\displaystyle\frac{di}{d\theta} & = & -3 \frac{J_2 A}{\Delta} \left(1 + J_2 X\cos(\theta) + J_2 Y\sin(\theta)\right) \sin(i)\cos(i)\sin(\theta)\cos(\theta); \nonumber \\
	\displaystyle\frac{d\Omega}{d\theta} & = & -3 \frac{J_2 A}{\Delta} \left(1 + J_2 X\cos(\theta) + J_2 Y\sin(\theta)\right) \cos(i) \sin^2(\theta); \nonumber \\
	\displaystyle\frac{dt}{d\theta} & = &  \sqrt[\leftroot{-1}\uproot{2}\scriptstyle 4]{\frac{R^6}{\mu^2 A^3}}\frac{1}{\Delta \left(1 + J_2 X\cos(\theta) + J_2 Y\sin(\theta)\right)^2},
\end{eqnarray}
where:
\begin{equation}
    \Delta = 1 + 3 J_2 A \left(1 + J_2 X \cos(\theta) + J_2 Y \sin(\theta)\right)\cos^2(i) \sin^2(\theta).
\end{equation}
It is important to note that in these expressions, the derivatives of both $X$ and $Y$ are no longer of order of magnitude $J_2$ and thus, a variation in the approach from Ref.~\cite{meanj2} is required to solve this difference. Particularly, each orbital element is approximated by a power series in the small parameter $J_2$ such that:
\begin{eqnarray}
    A & \approx & A_0 + A_1 J_2 + A_2 J_2^2, \nonumber \\
    X & \approx & X_{1} + X_{2} J_2, \nonumber \\
    Y & \approx & Y_{1} + Y_{2} J_2, \nonumber \\
    i & \approx & i_0 + i_1 J_2 + i_2 J_2^2, \nonumber \\
    \Omega & \approx & \Omega_0 + \Omega_1 J_2 + \Omega_2 J_2^2, \nonumber \\
    t & \approx & t_0 + t_{1} J_2 + t_{2} J_2^2,
\end{eqnarray}
where $X$ and $Y$ have only been expanded to the first power of $J_2$. The reason for that comes from the fact that $X_2$ and $Y_2$ are already representing the second order solution in the two components of the eccentricity vector. This approximation can be introduced in the differential equations to obtain the following expressions for zero order:
\begin{eqnarray}
    \displaystyle\frac{dA_0}{d\theta} & = & 0; \nonumber \\
    \displaystyle\frac{di_0}{d\theta} & = & 0; \nonumber \\
    \displaystyle\frac{d\Omega_0}{d\theta} & = & 0; \nonumber \\
    \displaystyle\frac{d t_0}{d\theta} & = & \sqrt[4]{\frac{R^6}{A_0^3 \mu^2}};
\end{eqnarray}
first order:
\begin{eqnarray}
    \displaystyle\frac{dA_1}{d\theta} & = & 6 A_0^2 \sin^2(i_0) \sin(2 \theta); \nonumber \\
    \displaystyle\frac{dX_1}{d\theta} & = & \frac{3}{2} A_0 \sin(\theta) (-1 - 4 \cos^2(\theta) \sin^2(i_0) + 3 \sin^2(i_0) \sin^2(\theta)); \nonumber \\
    \displaystyle\frac{dY_1}{d\theta} & = & \frac{3}{2} A_0 \cos(\theta) (1 - 7 \sin^2(i_0) \sin^2(\theta)); \nonumber \\
    \displaystyle\frac{di_1}{d\theta} & = & -3 A_0 \cos(i_0) \cos(\theta) \sin(i_0) \sin(\theta); \nonumber \\
    \displaystyle\frac{d\Omega_1}{d\theta} & = & -3 A_0 \cos(i_0) \sin^2(\theta); \nonumber \\
    \displaystyle\frac{d t_1}{d\theta} & = & \sqrt[4]{\frac{R^6}{A_0^3 \mu^2}} \Big((-\frac{3}{4}\frac{A_1}{A_0} - 3 A_0  \cos^2(i_0) \sin^2(\theta) +  (-2 X_0 \cos(\theta) - 2 Y_0 \sin(\theta))) \Big)
\end{eqnarray}
and second order:
\begin{eqnarray} \label{eq:difeq2}
    \displaystyle\frac{dA_2}{d\theta} & = & 12 A_0 \cos(\theta) \sin(i_0) \sin(\theta) (2 A_0 A_1 \cos(i_0) - 3 A_0^2 \cos^2(i_0) \sin(i_0) \sin^2(\theta) \nonumber \\
    & + & \sin(i_0) (2 i_1 + A_0 X_0 \cos(\theta) + A_0 Y_0 \sin(\theta))); \nonumber \\
    \displaystyle\frac{dX_2}{d\theta} & = & -\frac{3}{2} \sin(\theta) (i_1 + 5 A_0 X_0 \cos^3(\theta) \sin^2(i_0) + A_0 Y_0 (3 + \cos(2 i_0)) \sin(\theta) \nonumber \\
    & - & 3 (A_0^2 \cos^2(i_0) + i_1 \sin^2(i_0) + A_0 A_1 \sin(2 i_0)) \sin^2(\theta) \nonumber \\
    & - & 6 A_0 Y_0 \sin^2(i_0) \sin^3(\theta) + 9 A_0^2 \cos^2(i_0) \sin^2(i_0) \sin^4(\theta) \nonumber \\
    & + & 2 \cos^2(\theta) \sin(i_0) (4 A_0 A_1 \cos(i_0) + \sin(i_0) (2 i_1 + 3 A_0 Y_0 \sin(\theta))) \nonumber \\
    & + & A_0 X_0 \cos(\theta) (2 + \sin^2(i_0) (3 - 7 \sin^2(\theta))) - \frac{3}{4} A_0^2 \sin^2(2 i_0) \sin^2(2 \theta)); \nonumber \\
    \displaystyle\frac{dY_2}{d\theta} & = & \frac{3}{2} (2 A_0 X_0 \cos^2(\theta) - 2 A_0 Y_0 \cos^3(\theta) \sin^2(i_0) \sin(\theta) + \cos(\theta) (i_1 \nonumber \\
    & - & 7 A_0 A_1 \sin(2 i_0) \sin^2(\theta) - 7 \sin^2(i_0) \sin^2(\theta) (i_1 + 2 A_0 Y_0 \sin(\theta)) \nonumber \\
    & + & 3 A_0^2 \cos^2(i_0) \sin^2(\theta) (-1 + 7 \sin^2(i_0) \sin^2(\theta))) + A_0 (2 X_0 \cos^2(i_0) \sin^2(\theta) \nonumber \\
    & + & \sin(2 \theta) (Y_0 - 3 X_0 \sin^2(i_0) \sin(2 \theta)))); \nonumber \\
    \displaystyle\frac{di_2}{d\theta} & = & -3 \cos(\theta) \sin(\theta) (A_0 A_1 \cos^2(i_0) - A_0 A_1 \sin^2(i_0) - 3 A_0^2 \cos^3(i_0) \sin(i_0) \sin^2(\theta) \nonumber \\
    & + & \cos(i_0) \sin(i_0) (i_1 + A_0 X_0 \cos(\theta) + A_0 Y_0 \sin(\theta))); \nonumber \\
    \displaystyle\frac{d\Omega}{d\theta} & = & -3 \sin(\theta)^2 (-A_0 A_1 \sin(i_0) - 3 A_0^2 \cos^3(i_0) \sin^2(\theta) \nonumber \\
    & + & \cos(i_0) (i_1 + A_0 X_0 \cos(\theta) + A_0 Y_0 \sin(\theta))); \nonumber \\
    \displaystyle\frac{d t_2}{d\theta} & = & \sqrt[4]{\frac{R^6}{A_0^3 \mu^2}} \Big((\frac{1}{32 A_0^2}  (21 A_1^2 - 24 A_0 A_2 + 48 A_0 A_1 X_0 \cos(\theta) - 64 A_0^2 X_1 \cos(\theta) \nonumber \\
    & + & 96 A_0^2 X_0^2 \cos^2(\theta) + 48 A_0 A_1 Y_0 \sin(\theta) - 64 A_0^2 Y_1 \sin(\theta) \nonumber \\
    & + & 192 A_0^2 X_0 Y_0 \cos(\theta) \sin(\theta) + 96 A_0^2 Y_0^2 \sin^2(\theta)   - 24 A_0^2 A_1 \cos(i_0)^2 \sin^2(\theta) \nonumber \\
    & + & 96 A_0^3 X_0 \cos^2(i_0) \cos(\theta) \sin^2(\theta) + 
    192 A_0^3 i_1 \cos(i_0) \sin(i_0) \sin^2(\theta) \nonumber \\
    & + & 96 A_0^3 Y_0 \cos(i_0)^2 \sin^3(\theta) + 
    288 A_0^4 \cos^4(i_0) \sin^4(\theta))) \Big).
\end{eqnarray}
The following sections provide the solutions to these differential equations, as well as the transformation from osculating to mean elements, and the secular variation of these variables. These results are then used to find the frozen conditions of the orbit and the perturbed period of the orbits under $J_2$.


\section{Approximate analytic solution} \label{sec:solution}

The approximate analytic solution can be obtained by a direct integration from the approximate differential equations presented in the previous section. Particularly, let $phi$ be one of the orbital elements considered (including time). Then, $\phi_n$ is the $n$th-order solution on that variable, which can be obtained using the integral:
\begin{equation}
    \phi_n = \int_{\theta_0}^{\theta} \displaystyle\frac{d \phi_n}{d\theta},
\end{equation}
where $\theta_0$ is the initial value of the argument of latitude. This means that the evolution of the variable is the result of the sum of the power series derived, particularly:
\begin{equation}
    \phi = \sum_{n=0}^{m} \phi_n J_2^n.
\end{equation}
if a solution of order $m$ is considered. The following subsections include the results for each individual order.

\subsection{Zero order solution}

In the zero order solution, we have that $A_0 = A(\theta = \theta_0)$, $i_0 = i(\theta = \theta_0)$, and $\Omega_0 = \Omega(\theta = \theta_0)$ since the derivatives of these variables are zero. On the other hand, the time evolution is provided by:
\begin{equation}
    t_0 = \sqrt[4]{\frac{R^6}{A_0^3 \mu^2}} (\theta - \theta_0);
\end{equation}
if we consider the initial time of the propagation to be zero.

\subsection{First order solution}

The initial conditions for the first order solution are $A_1 = i_1 = \Omega_1 = t_1 = 0 $ since they represent the first order deviation from the unperturbed problem. In contrast, two normalized components of the eccentricity have an initial condition dependent on the initial value of $e_x$ and $e_y$ since $X$ and $Y$ do not have differential equation for order zero. That is, their initial conditions are:
\begin{eqnarray}
    X_1 (t=0) = X_0 = \displaystyle\frac{e_{x}(t = 0)}{J_2}, &  & Y_1 (t=0) = Y_0 = \displaystyle\frac{e_{y}(t = 0)}{J_2},
\end{eqnarray}
respectively. That way, the first order solution of the problem is:
\begin{eqnarray} 
    A_1 & = & 3 A_0^2 (\cos(2 \theta_0) - \cos(2 \theta)) \sin^2(i_0); \nonumber \\
    X_1 & = & X_0 + \displaystyle\frac{1}{8} (12 A_0 (-\cos(\theta_0) + \cos(\theta)) + 
   A_0 (15 \cos(\theta_0) - 7 \cos(3 \theta_0) \nonumber \\
   & - & 15 \cos(\theta) + 7 \cos(3 \theta)) \sin^2(i_0)); \nonumber \\
   Y_1 & = & Y_0 + \displaystyle\frac{1}{2} A_0 (3 (-\sin(\theta_0) + \sin(\theta)) + 7 \sin^2(i_0) (\sin^3(\theta_0) - \sin^3(\theta))); \nonumber \\
   i_1 & = & -\displaystyle\frac{3}{8} A_0 (\cos(2 \theta_0) - \cos(2 \theta)) \sin(2 i_0) \nonumber \\
   \Omega_1 & = & \displaystyle\frac{3}{2} A_0 \cos(i_0) (\theta_0 - \theta - \cos(\theta_0) \sin(\theta_0) + \cos(\theta) \sin(\theta)); \nonumber
\end{eqnarray}
\begin{eqnarray} \label{eq:1stsolution}
   t_1 & = & \displaystyle\frac{1}{4} \sqrt[4]{\frac{R^6}{A_0^3 \mu^2}} (12 A_0 \theta_0 - 12 A_0 \theta - 18 A_0 \theta_0 \sin^2(i_0) + 18 A_0 \theta \sin^2(i_0)  \nonumber \\
   & + & 8 X_0 \sin(\theta_0) + 4 \cos(\theta) (2 Y_0 - 3 A_0 \sin(\theta_0) + 7 A_0 \sin^2(i_0) \sin^3(\theta_0))  \nonumber \\
   & + & 16 A_0 \sin^2(i_0) \sin(2 \theta_0) - 7/2 A_0 \sin^2(i_0) \sin(4 \theta_0) - 8 X_0 \sin(\theta)  \nonumber \\
   & + & 7 A_0 \cos(3 \theta_0) \sin^2(i_0) \sin(\theta) + 6 A_0 \cos^2(i_0) (\theta_0 - \theta - \cos(\theta_0) \sin(\theta_0)  \nonumber \\
   & + & \cos(\theta) \sin(\theta)) + 9 A_0 \sin^2(i_0) ((\theta_0 - \theta) \cos(2 \theta_0) - \cos(\theta_0) \sin(\theta_0)  \nonumber \\
   & + & \cos(\theta) \sin(\theta)) - 
   \cos(\theta_0) (8 Y_0 - 12 A_0 \sin(\theta) + A_0 \sin^2(i_0) (28 \sin^3(\theta_0)  \nonumber \\
   & + & 15 \sin(\theta))) - 5 A_0 \sin^2(i_0) \sin(2 \theta)).
\end{eqnarray}

\subsection{Second order solution} \label{sec:2ndsolution}

The second order solution requires to introduce the results from the first order and second order solutions into Eq.~\eqref{eq:difeq2}. Once this is done, it is again possible to obtain the second order evolution of the orbital elements by a direct integration of the equations. Particularly, the solution for $A_2$ is:
\begin{eqnarray}
    A_2 & = & -\displaystyle\frac{1}{64} A_0^2 \sin^2(
  i_0) (12 A_0 (3 + 5 \cos(2 i_0)) \cos^4(\theta_0) - 64 X_0 \cos(3 \theta_0)  \nonumber \\
  & + & 
   192 X_0 \cos(\theta) + 108 A_0 \cos^4(\theta) + 180 A_0 \cos(2 i_0) \cos^4(\theta)  \nonumber \\
   & - & 
   48 \cos(\theta_0) (4 X_0 + A_0 (3 + 5 \cos(2 i_0)) \cos^3(\theta)) + 64 X_0 \cos(3 \theta)  \nonumber \\
   & + & 
   42 A_0 \cos(2 \theta_0) \sin^2(i_0) - 1152 A_0 \cos^2(2 \theta_0) \sin^2(i_0)  \nonumber \\
   & + & 
   224 A_0 \cos^3(\theta_0) \cos(3 \theta_0) \sin^2(i_0) + 267 A_0 \cos(4 \theta_0) \sin^2(i_0)  \nonumber \\
   & - & 
   14 A_0 \cos(6 \theta_0) \sin^2(i_0) - 224 A_0 \cos(3 \theta_0) \cos^3(\theta) \sin^2(i_0)  \nonumber \\
   & - & 
   42 A_0 \cos(2 \theta) \sin^2(i_0) + 1152 A_0 \cos(2 \theta_0) \cos(2 \theta) \sin^2(i_0)  \nonumber \\
   & - & 
   267 A_0 \cos(4 \theta) \sin^2(i_0) + 14 A_0 \cos(6 \theta) \sin^2(i_0) + 
   192 Y_0 \sin(\theta_0)  \nonumber \\
   & - & 96 A_0 \sin^4(\theta_0) + 448 A_0 \sin^2(i_0) \sin^6(\theta_0) - 
   64 Y_0 \sin(3 \theta_0) - 192 Y_0 \sin(\theta)  \nonumber \\
   & + & 
   576 A_0 \cos^2(i_0) (\cos(2 \theta_0) - \cos(2 \theta)) \sin^2(\theta) + 
   384 A_0 \sin(\theta_0) \sin^3(\theta)  \nonumber \\
   & - & 896 A_0 \sin^2(i_0) \sin^3(\theta_0) \sin^3(\theta) - 
   288 A_0 \sin^4(\theta) + 448 A_0 \sin^2(i_0) \sin^6(\theta)  \nonumber \\
   & + & 64 Y_0 \sin(3 \theta)), 
\end{eqnarray}
the solution for $X_2$ is:
\begin{eqnarray}
    X_2 & = & \displaystyle\frac{1}{2048} A_0 (2304 \theta_0 Y_0 - 2304 \theta Y_0 + 3840 (\theta_0 - \theta) Y_0 \cos(2 i_0) \nonumber \\
   & - & 
    156 A_0 \cos(2 i_0 - 5 \theta_0) + 81 A_0 \cos(4 i_0 - 5 \theta_0) - 
    472 A_0 \cos(2 i_0 - 3 \theta_0) \nonumber \\
   & + & 128 A_0 \cos(4 i_0 - 3 \theta_0) + 
    288 X_0 \cos(2 (i_0 - 2 \theta_0)) + 384 X_0 \cos(2 (i_0 - \theta_0)) \nonumber \\
   & - & 
    1884 A_0 \cos(2 i_0 - \theta_0) + 135 A_0 \cos(4 i_0 - \theta_0) - 2262 A_0 \cos(\theta_0) \nonumber \\
   & - & 
    2304 X_0 \cos(2 \theta_0) - 464 A_0 \cos(3 \theta_0) - 576 X_0 \cos(4 \theta_0) + 
    150 A_0 \cos(5 \theta_0) \nonumber \\
   & + & 384 X_0 \cos(2 (i_0 + \theta_0)) - 1884 A_0 \cos(2 i_0 + \theta_0) + 
    135 A_0 \cos(4 i_0 + \theta_0) \nonumber \\
   & + & 288 X_0 \cos(2 (i_0 + 2 \theta_0)) - 
    472 A_0 \cos(2 i_0 + 3 \theta_0) + 128 A_0 \cos(4 i_0 + 3 \theta_0) \nonumber \\
   & - & 
    156 A_0 \cos(2 i_0 + 5 \theta_0) + 81 A_0 \cos(4 i_0 + 5 \theta_0) + 
    48 A_0 \cos(2 i_0 - 5 \theta) \nonumber \\
   & - & 96 A_0 \cos(4 i_0 - 5 \theta) + 
    108 A_0 \cos(2 i_0 - \theta_0 - 4 \theta) - 27 A_0 \cos(4 i_0 - \theta_0 - 4 \theta) \nonumber \\
   & + & 
    108 A_0 \cos(\theta_0 - 4 \theta) - 216 A_0 \cos(2 i_0 + \theta_0 - 4 \theta) + 
    162 A_0 \cos(4 i_0 + \theta_0 - 4 \theta) \nonumber \\
   & - & 378 A_0 \cos(3 \theta_0 - 4 \theta) + 
    252 A_0 \cos(2 i_0 + 3 \theta_0 - 4 \theta) \nonumber
\end{eqnarray}
\begin{eqnarray}
   & - & 63 A_0 \cos(4 i_0 + 3 \theta_0 - 4 \theta) + 
    688 A_0 \cos(2 i_0 - 3 \theta) - 344 A_0 \cos(4 i_0 - 3 \theta) \nonumber \\
   & - & 
    672 A_0 \cos(2 i_0 - 2 \theta_0 - 3 \theta) + 252 A_0 \cos(4 i_0 - 2 \theta_0 - 3 \theta) \nonumber \\
   & + & 
    840 A_0 \cos(2 \theta_0 - 3 \theta) - 672 A_0 \cos(2 i_0 + 2 \theta_0 - 3 \theta) \nonumber \\
   & + &
    252 A_0 \cos(4 i_0 + 2 \theta_0 - 3 \theta) - 288 X_0 \cos(2 (i_0 - 2 \theta)) \nonumber \\
   & + & 
    672 A_0 \cos(2 i_0 - 3 \theta_0 - 2 \theta) - 210 A_0 \cos(4 i_0 - 3 \theta_0 - 2 \theta) \nonumber \\
   & - & 
    288 A_0 \cos(2 i_0 - \theta_0 - 2 \theta) + 594 A_0 \cos(4 i_0 - \theta_0 - 2 \theta) \nonumber \\
   & - & 
    1476 A_0 \cos(\theta_0 - 2 \theta) - 576 A_0 \cos(2 i_0 + \theta_0 - 2 \theta) - 
    414 A_0 \cos(4 i_0 + \theta_0 - 2 \theta) \nonumber \\
   & - & 252 A_0 \cos(3 \theta_0 - 2 \theta) + 
    126 A_0 \cos(4 i_0 + 3 \theta_0 - 2 \theta) - 384 X_0 \cos(2 (i_0 - \theta)) \nonumber \\
   & + & 
    2592 A_0 \cos(2 i_0 - \theta) + 504 A_0 \cos(4 i_0 - \theta) + 
    288 A_0 \cos(2 i_0 - 2 \theta_0 - \theta) \nonumber \\
   & - & 540 A_0 \cos(4 i_0 - 2 \theta_0 - \theta) + 
    504 A_0 \cos(2 \theta_0 - \theta) + 288 A_0 \cos(2 i_0 + 2 \theta_0 - \theta) \nonumber \\
   & - & 
    540 A_0 \cos(4 i_0 + 2 \theta_0 - \theta) + 3024 A_0 \cos(\theta) + 2304 X_0 \cos(2 \theta) - 
    688 A_0 \cos(3 \theta) \nonumber \\
   & + & 576 X_0 \cos(4 \theta) + 96 A_0 \cos(5 \theta) - 
    384 X_0 \cos(2 (i_0 + \theta)) + 2592 A_0 \cos(2 i_0 + \theta) \nonumber \\
   & + & 
    504 A_0 \cos(4 i_0 + \theta) + 288 A_0 \cos(2 i_0 - 2 \theta_0 + \theta) - 
    540 A_0 \cos(4 i_0 - 2 \theta_0 + \theta) \nonumber \\
   & + & 504 A_0 \cos(2 \theta_0 + \theta) + 
    288 A_0 \cos(2 i_0 + 2 \theta_0 + \theta) - 540 A_0 \cos(4 i_0 + 2 \theta_0 + \theta) \nonumber \\
   & - & 
    288 X_0 \cos(2 (i_0 + 2 \theta)) + 126 A_0 \cos(4 i_0 - 3 \theta_0 + 2 \theta) \nonumber \\
   & - & 
    576 A_0 \cos(2 i_0 - \theta_0 + 2 \theta) - 414 A_0 \cos(4 i_0 - \theta_0 + 2 \theta) + 
    540 A_0 \cos(\theta_0 + 2 \theta) \nonumber \\
   & - & 288 A_0 \cos(2 i_0 + \theta_0 + 2 \theta) + 
    594 A_0 \cos(4 i_0 + \theta_0 + 2 \theta) - 924 A_0 \cos(3 \theta_0 + 2 \theta) \nonumber \\
   & + & 
    672 A_0 \cos(2 i_0 + 3 \theta_0 + 2 \theta) - 210 A_0 \cos(4 i_0 + 3 \theta_0 + 2 \theta) + 
    688 A_0 \cos(2 i_0 + 3 \theta) \nonumber \\
   & - & 344 A_0 \cos(4 i_0 + 3 \theta) - 
    672 A_0 \cos(2 i_0 - 2 \theta_0 + 3 \theta) + 252 A_0 \cos(4 i_0 - 2 \theta_0 + 3 \theta) \nonumber \\
   & + & 
    840 A_0 \cos(2 \theta_0 + 3 \theta) - 672 A_0 \cos(2 i_0 + 2 \theta_0 + 3 \theta) + 
    252 A_0 \cos(4 i_0 + 2 \theta_0 + 3 \theta) \nonumber \\
   & + & 252 A_0 \cos(2 i_0 - 3 \theta_0 + 4 \theta) - 
    63 A_0 \cos(4 i_0 - 3 \theta_0 + 4 \theta) \nonumber \\
   & - & 216 A_0 \cos(2 i_0 - \theta_0 + 4 \theta) + 
    162 A_0 \cos(4 i_0 - \theta_0 + 4 \theta) - 162 A_0 \cos(\theta_0 + 4 \theta) \nonumber \\
   & + & 
    108 A_0 \cos(2 i_0 + \theta_0 + 4 \theta) - 27 A_0 \cos(4 i_0 + \theta_0 + 4 \theta) + 
    48 A_0 \cos(2 i_0 + 5 \theta) \nonumber \\
   & - & 96 A_0 \cos(4 i_0 + 5 \theta) + 
    336 A_0 \theta_0 \sin(2 i_0 - 3 \theta_0) - 336 A_0 \theta \sin(2 i_0 - 3 \theta_0) \nonumber \\
   & - & 
    420 A_0 \theta_0 \sin(4 i_0 - 3 \theta_0) + 420 A_0 \theta \sin(4 i_0 - 3 \theta_0) - 
    288 Y_0 \sin(2 (i_0 - 2 \theta_0)) \nonumber \\
   & + & 1536 Y_0 \sin(2 (i_0 - \theta_0)) + 
    1872 A_0 \theta_0 \sin(2 i_0 - \theta_0) - 1872 A_0 \theta \sin(2 i_0 - \theta_0) \nonumber \\
   & + & 
    1260 A_0 \theta_0 \sin(4 i_0 - \theta_0) - 1260 A_0 \theta \sin(4 i_0 - \theta_0) - 
    2952 A_0 \theta_0 \sin(\theta_0) \nonumber \\
   & + & 2952 A_0 \theta \sin(\theta_0) - 168 A_0 \theta_0 \sin(3 \theta_0) + 
    168 A_0 \theta \sin(3 \theta_0) - 576 Y_0 \sin(4 \theta_0) \nonumber \\
   & - & 1536 Y_0 \sin(2 (i_0 + \theta_0)) - 
    1872 A_0 \theta_0 \sin(2 i_0 + \theta_0) + 1872 A_0 \theta \sin(2 i_0 + \theta_0) \nonumber \\
   & - & 
    1260 A_0 \theta_0 \sin(4 i_0 + \theta_0) + 1260 A_0 \theta \sin(4 i_0 + \theta_0) + 
    288 Y_0 \sin(2 (i_0 + 2 \theta_0)) \nonumber \\
   & - & 336 A_0 \theta_0 \sin(2 i_0 + 3 \theta_0) + 
    336 A_0 \theta \sin(2 i_0 + 3 \theta_0) + 420 A_0 \theta_0 \sin(4 i_0 + 3 \theta_0) \nonumber \\
   & - & 
    420 A_0 \theta \sin(4 i_0 + 3 \theta_0) + 288 Y_0 \sin(2 (i_0 - 2 \theta)) - 
    1536 Y_0 \sin(2 (i_0 - \theta)) \nonumber \\
   & + & 576 Y_0 \sin(4 \theta) + 
    1536 Y_0 \sin(2 (i_0 + \theta)) - 288 Y_0 \sin(2 (i_0 + 2 \theta))),
\end{eqnarray}
the solution for $Y_2$ is:
\begin{eqnarray}
Y_2 & = & \displaystyle\frac{1}{2048} A_0 (-2304 \theta_0 X_0 + 2304 \theta X_0 - 3840 (\theta_0 - \theta) X_0 \cos(2 i_0)   \nonumber \\
& + & 
    336 A_0 (\theta_0 - \theta) \cos(2 i_0 - 3 \theta_0) - 420 A_0 \theta_0 \cos(4 i_0 - 3 \theta_0)   \nonumber \\
    & + & 
    420 A_0 \theta \cos(4 i_0 - 3 \theta_0) - 288 Y_0 \cos(2 (i_0 - 2 \theta_0)) + 
    1536 Y_0 \cos(2 (i_0 - \theta_0))   \nonumber \\
    & + & 2160 A_0 \theta_0 \cos(2 i_0 - \theta_0) - 
    2160 A_0 \theta \cos(2 i_0 - \theta_0) + 900 A_0 \theta_0 \cos(4 i_0 - \theta_0)   \nonumber
\end{eqnarray}
\begin{eqnarray}
    & - & 
    900 A_0 \theta \cos(4 i_0 - \theta_0) + 3096 A_0 \theta_0 \cos(\theta_0) - 3096 A_0 \theta \cos(\theta_0)   \nonumber \\
    & - & 
    1536 Y_0 \cos(2 \theta_0) + 168 A_0 \theta_0 \cos(3 \theta_0) - 168 A_0 \theta \cos(3 \theta_0) + 
    576 Y_0 \cos(4 \theta_0)   \nonumber \\
    & + & 1536 Y_0 \cos(2 (i_0 + \theta_0)) + 
    2160 A_0 \theta_0 \cos(2 i_0 + \theta_0) - 2160 A_0 \theta \cos(2 i_0 + \theta_0)   \nonumber \\
    & + & 
    900 A_0 \theta_0 \cos(4 i_0 + \theta_0) - 900 A_0 \theta \cos(4 i_0 + \theta_0) - 
    288 Y_0 \cos(2 (i_0 + 2 \theta_0))   \nonumber \\
    & + & 336 A_0 \theta_0 \cos(2 i_0 + 3 \theta_0) - 
    336 A_0 \theta \cos(2 i_0 + 3 \theta_0) - 420 A_0 \theta_0 \cos(4 i_0 + 3 \theta_0)   \nonumber \\
    & + & 
    420 A_0 \theta \cos(4 i_0 + 3 \theta_0) + 288 Y_0 \cos(2 (i_0 - 2 \theta)) - 
    1536 Y_0 \cos(2 (i_0 - \theta))   \nonumber \\
    & + & 1536 Y_0 \cos(2 \theta) - 576 Y_0 \cos(4 \theta) - 
    1536 Y_0 \cos(2 (i_0 + \theta))   \nonumber \\
    & + & 288 Y_0 \cos(2 (i_0 + 2 \theta)) + 
    156 A_0 \sin(2 i_0 - 5 \theta_0) - 81 A_0 \sin(4 i_0 - 5 \theta_0)   \nonumber \\
    & + & 
    808 A_0 \sin(2 i_0 - 3 \theta_0) - 248 A_0 \sin(4 i_0 - 3 \theta_0) - 
    288 X_0 \sin(2 (i_0 - 2 \theta_0))   \nonumber \\
    & - & 384 X_0 \sin(2 (i_0 - \theta_0)) + 
    1164 A_0 \sin(2 i_0 - \theta_0) + 1629 A_0 \sin(4 i_0 - \theta_0)   \nonumber \\
    & - & 2478 A_0 \sin(\theta_0) - 
    768 X_0 \sin(2 \theta_0) - 32 A_0 \sin(3 \theta_0) - 576 X_0 \sin(4 \theta_0)   \nonumber \\
    & + & 
    150 A_0 \sin(5 \theta_0) + 384 X_0 \sin(2 (i_0 + \theta_0)) - 1164 A_0 \sin(2 i_0 + \theta_0)   \nonumber \\
    & - & 
    1629 A_0 \sin(4 i_0 + \theta_0) + 288 X_0 \sin(2 (i_0 + 2 \theta_0)) - 
    808 A_0 \sin(2 i_0 + 3 \theta_0)   \nonumber \\
    & + & 248 A_0 \sin(4 i_0 + 3 \theta_0) - 
    156 A_0 \sin(2 i_0 + 5 \theta_0) + 81 A_0 \sin(4 i_0 + 5 \theta_0)   \nonumber \\
    & - & 
    48 A_0 \sin(2 i_0 - 5 \theta) + 96 A_0 \sin(4 i_0 - 5 \theta) - 
    108 A_0 \sin(2 i_0 - \theta_0 - 4 \theta)   \nonumber \\
    & + & 27 A_0 \sin(4 i_0 - \theta_0 - 4 \theta) - 
    108 A_0 \sin(\theta_0 - 4 \theta) + 216 A_0 \sin(2 i_0 + \theta_0 - 4 \theta)   \nonumber \\
    & - & 
    162 A_0 \sin(4 i_0 + \theta_0 - 4 \theta) + 378 A_0 \sin(3 \theta_0 - 4 \theta) - 
    252 A_0 \sin(2 i_0 + 3 \theta_0 - 4 \theta)   \nonumber \\
    & + & 63 A_0 \sin(4 i_0 + 3 \theta_0 - 4 \theta) - 
    592 A_0 \sin(2 i_0 - 3 \theta) + 248 A_0 \sin(4 i_0 - 3 \theta)   \nonumber \\
    & + & 
    672 A_0 \sin(2 i_0 - 2 \theta_0 - 3 \theta) - 252 A_0 \sin(4 i_0 - 2 \theta_0 - 3 \theta) - 
    840 A_0 \sin(2 \theta_0 - 3 \theta)   \nonumber \\
    & + & 672 A_0 \sin(2 i_0 + 2 \theta_0 - 3 \theta) - 
    252 A_0 \sin(4 i_0 + 2 \theta_0 - 3 \theta) + 288 X_0 \sin(2 (i_0 - 2 \theta))   \nonumber \\
    & - & 
    672 A_0 \sin(2 i_0 - 3 \theta_0 - 2 \theta) + 210 A_0 \sin(4 i_0 - 3 \theta_0 - 2 \theta)   \nonumber \\
    & + & 
    432 A_0 \sin(2 i_0 - \theta_0 - 2 \theta) - 594 A_0 \sin(4 i_0 - \theta_0 - 2 \theta) + 
    900 A_0 \sin(\theta_0 - 2 \theta)   \nonumber \\
    & - & 288 A_0 \sin(2 i_0 + \theta_0 - 2 \theta) + 
    414 A_0 \sin(4 i_0 + \theta_0 - 2 \theta) - 420 A_0 \sin(3 \theta_0 - 2 \theta)   \nonumber \\
    & + & 
    336 A_0 \sin(2 i_0 + 3 \theta_0 - 2 \theta) - 126 A_0 \sin(4 i_0 + 3 \theta_0 - 2 \theta) + 
    384 X_0 \sin(2 (i_0 - \theta))   \nonumber \\
    & - & 1824 A_0 \sin(2 i_0 - \theta) - 
    1224 A_0 \sin(4 i_0 - \theta) - 864 A_0 \sin(2 i_0 - 2 \theta_0 - \theta)   \nonumber \\
    & + & 
    756 A_0 \sin(4 i_0 - 2 \theta_0 - \theta) + 216 A_0 \sin(2 \theta_0 - \theta) - 
    864 A_0 \sin(2 i_0 + 2 \theta_0 - \theta) \nonumber \\
& + & 756 A_0 \sin(4 i_0 + 2 \theta_0 - \theta) + 
    3120 A_0 \sin(\theta) + 768 X_0 \sin(2 \theta) - 688 A_0 \sin(3 \theta) \nonumber \\
& + & 
    576 X_0 \sin(4 \theta) + 96 A_0 \sin(5 \theta) - 384 X_0 \sin(2 (i_0 + \theta)) + 
    1824 A_0 \sin(2 i_0 + \theta) \nonumber \\
& + & 1224 A_0 \sin(4 i_0 + \theta) + 
    864 A_0 \sin(2 i_0 - 2 \theta_0 + \theta) - 756 A_0 \sin(4 i_0 - 2 \theta_0 + \theta) \nonumber \\
& - & 
    216 A_0 \sin(2 \theta_0 + \theta) + 864 A_0 \sin(2 i_0 + 2 \theta_0 + \theta) - 
    756 A_0 \sin(4 i_0 + 2 \theta_0 + \theta) \nonumber \\
& - & 288 X_0 \sin(2 (i_0 + 2 \theta)) - 
    336 A_0 \sin(2 i_0 - 3 \theta_0 + 2 \theta) \nonumber \\
& + & 126 A_0 \sin(4 i_0 - 3 \theta_0 + 2 \theta) + 
    288 A_0 \sin(2 i_0 - \theta_0 + 2 \theta) \nonumber \\
& - & 414 A_0 \sin(4 i_0 - \theta_0 + 2 \theta) + 
    828 A_0 \sin(\theta_0 + 2 \theta) - 432 A_0 \sin(2 i_0 + \theta_0 + 2 \theta) \nonumber \\
& + &
    594 A_0 \sin(4 i_0 + \theta_0 + 2 \theta) - 924 A_0 \sin(3 \theta_0 + 2 \theta) + 
    672 A_0 \sin(2 i_0 + 3 \theta_0 + 2 \theta) \nonumber \\
& - & 210 A_0 \sin(4 i_0 + 3 \theta_0 + 2 \theta) + 
    592 A_0 \sin(2 i_0 + 3 \theta) - 248 A_0 \sin(4 i_0 + 3 \theta) \nonumber \\
& - &
    672 A_0 \sin(2 i_0 - 2 \theta_0 + 3 \theta) + 252 A_0 \sin(4 i_0 - 2 \theta_0 + 3 \theta) + 
    840 A_0 \sin(2 \theta_0 + 3 \theta) \nonumber \\
& - & 672 A_0 \sin(2 i_0 + 2 \theta_0 + 3 \theta) + 
    252 A_0 \sin(4 i_0 + 2 \theta_0 + 3 \theta) \nonumber \\
& + & 252 A_0 \sin(2 i_0 - 3 \theta_0 + 4 \theta) - 
    63 A_0 \sin(4 i_0 - 3 \theta_0 + 4 \theta) \nonumber \\
& - & 216 A_0 \sin(2 i_0 - \theta_0 + 4 \theta) + 
    162 A_0 \sin(4 i_0 - \theta_0 + 4 \theta) - 162 A_0 \sin(\theta_0 + 4 \theta) \nonumber
\end{eqnarray}
\begin{eqnarray}
& + & 
    108 A_0 \sin(2 i_0 + \theta_0 + 4 \theta) - 27 A_0 \sin(4 i_0 + \theta_0 + 4 \theta) \qquad\qquad\qquad\qquad\qquad \nonumber \\
& + &
    48 A_0 \sin(2 i_0 + 5 \theta) - 96 A_0 \sin(4 i_0 + 5 \theta)),
\end{eqnarray}
the solution for $i_2$ is:
\begin{eqnarray}
i_2 & = & -\displaystyle\frac{1}{256} A_0 \cos(i_0) \sin(
  i_0) (-12 A_0 (3 + 5 \cos(2 i_0)) \cos^4(\theta_0) + 64 X_0 \cos(3 \theta_0)   \nonumber \\
  & - & 
   192 X_0 \cos(\theta) - 108 A_0 \cos^4(\theta) - 180 A_0 \cos(2 i_0) \cos^4(\theta) - 
   64 X_0 \cos(3 \theta)   \nonumber \\
   & + & 
   36 A_0 \cos^2(
     i_0) (-2 + \cos(4 \theta_0) + 4 \cos(2 \theta_0) (-2 + \cos(2 \theta)) + 
      8 \cos(2 \theta)   \nonumber \\
      & - & 3 \cos(4 \theta)) - 42 A_0 \cos(2 \theta_0) \sin^2(i_0) + 
   720 A_0 \cos^2(2 \theta_0) \sin^2(i_0)   \nonumber \\
   & - & 56 A_0 \cos^2(3 \theta_0) \sin^2(i_0) - 
   159 A_0 \cos(4 \theta_0) \sin^2(i_0) + 14 A_0 \cos(6 \theta_0) \sin^2(i_0)   \nonumber \\
   & + & 
   168 A_0 \cos(3 \theta_0) \cos(\theta) \sin^2(i_0) + 42 A_0 \cos(2 \theta) \sin^2(i_0)   \nonumber \\
   & - & 
   720 A_0 \cos(2 \theta_0) \cos(2 \theta) \sin^2(i_0) + 
   56 A_0 \cos(3 \theta_0) \cos(3 \theta) \sin^2(i_0)   \nonumber \\
   & + & 159 A_0 \cos(4 \theta) \sin^2(i_0) - 
   14 A_0 \cos(6 \theta) \sin^2(i_0) + 
   24 \cos(\theta_0) (8 X_0   \nonumber \\
   & + & 2 A_0 (3 + 5 \cos(2 i_0)) \cos^3(\theta) - 
      7 A_0 \cos(3 \theta_0) \sin^2(i_0)) - 192 Y_0 \sin(\theta_0)   \nonumber \\
      & + & 96 A_0 \sin^4(\theta_0) - 
   448 A_0 \sin^2(i_0) \sin^6(\theta_0) + 64 Y_0 \sin(3 \theta_0) + 192 Y_0 \sin(\theta)   \nonumber \\
   & - & 
   384 A_0 \sin(\theta_0) \sin^3(\theta) + 896 A_0 \sin^2(i_0) \sin^3(\theta_0) \sin^3(\theta) + 
   288 A_0 \sin^4(\theta)   \nonumber \\
   & - & 448 A_0 \sin^2(i_0) \sin^6(\theta) - 64 Y_0 \sin(3 \theta)), \nonumber 
\end{eqnarray}
the solution for $\Omega_2$ is:
\begin{eqnarray}
\Omega_2 & = & - \displaystyle\frac{1}{128} A_0 \cos(
  i_0) (24 A_0 \theta_0 - 24 A_0 \theta + 120 A_0 (\theta_0 - \theta) \cos(2 i_0) \nonumber \\
   & + &  
   180 A_0 (\theta_0 - \theta) \cos(2 (i_0 - \theta_0)) + 288 Y_0 \cos(\theta_0) - 
   360 A_0 \theta_0 \cos(2 \theta_0) \nonumber \\
   & + &  360 A_0 \theta \cos(2 \theta_0) - 32 Y_0 \cos(3 \theta_0) + 
   180 A_0 \theta_0 \cos(2 (i_0 + \theta_0))\nonumber \\
   & - &  180 A_0 \theta \cos(2 (i_0 + \theta_0)) - 
   288 Y_0 \cos(\theta) + 32 Y_0 \cos(3 \theta) \nonumber \\
   & - &  3 A_0 \sin(2 (i_0 - 2 \theta_0)) + 
   126 A_0 \sin(2 (i_0 - \theta_0)) - 96 X_0 \sin(\theta_0) \nonumber \\
   & - & 84 A_0 \sin(2 \theta_0) + 
   32 X_0 \sin(3 \theta_0) + 30 A_0 \sin(4 \theta_0) - 126 A_0 \sin(2 (i_0 + \theta_0)) \nonumber \\
   & + &  
   3 A_0 \sin(2 (i_0 + 2 \theta_0)) + 3 A_0 \sin(2 i_0 - \theta_0 - 3 \theta) - 
   12 A_0 \sin(\theta_0 - 3 \theta) \nonumber \\
   & - &  18 A_0 \sin(2 i_0 + \theta_0 - 3 \theta) + 
   7 A_0 \sin(2 i_0 + 3 \theta_0 - 3 \theta) + 24 A_0 \sin(2 (i_0 - 2 \theta)) \nonumber \\
   & - &  
   78 A_0 \sin(2 (i_0 - \theta)) + 21 A_0 \sin(2 i_0 - 3 \theta_0 - \theta) - 
   45 A_0 \sin(2 (i_0 - \theta_0 - \theta)) \nonumber \\
   & - &  72 A_0 \sin(2 i_0 - \theta_0 - \theta) + 
   54 A_0 \sin(\theta_0 - \theta) + 90 A_0 \sin(2 (\theta_0 - \theta)) \nonumber \\
   & - &  
   14 A_0 \sin(3 (\theta_0 - \theta)) - 45 A_0 \sin(2 (i_0 + \theta_0 - \theta)) + 
   117 A_0 \sin(2 i_0 + \theta_0 - \theta) \nonumber \\
   & + &  84 A_0 \sin(3 \theta_0 - \theta) - 
   42 A_0 \sin(2 i_0 + 3 \theta_0 - \theta) + 96 X_0 \sin(\theta) - 12 A_0 \sin(2 \theta) \nonumber \\
   & - &  
   32 X_0 \sin(3 \theta) + 12 A_0 \sin(4 \theta) + 78 A_0 \sin(2 (i_0 + \theta)) + 
   42 A_0 \sin(2 i_0 - 3 \theta_0 + \theta) \nonumber \\
   & + &  45 A_0 \sin(2 (i_0 - \theta_0 + \theta)) - 
   117 A_0 \sin(2 i_0 - \theta_0 + \theta) - 90 A_0 \sin(2 (\theta_0 + \theta)) \nonumber \\
   & + &  
   45 A_0 \sin(2 (i_0 + \theta_0 + \theta)) + 72 A_0 \sin(2 i_0 + \theta_0 + \theta) + 
   42 A_0 \sin(3 \theta_0 + \theta) \nonumber \\
   & - &  21 A_0 \sin(2 i_0 + 3 \theta_0 + \theta) - 
   24 A_0 \sin(2 (i_0 + 2 \theta)) - 7 A_0 \sin(2 i_0 - 3 \theta_0 + 3 \theta) \nonumber \\
   & + &  
   18 A_0 \sin(2 i_0 - \theta_0 + 3 \theta) + 6 A_0 \sin(\theta_0 + 3 \theta) - 
   3 A_0 \sin(2 i_0 + \theta_0 + 3 \theta)),
\end{eqnarray}
and finally, the solution for $t_2$ is:
\begin{eqnarray}
    t_2 & = & -\frac{1}{4096}\sqrt[4]{\frac{R^6}{A_0^3 \mu^2}} (11928 \theta_0
   A_0^2-11928 \theta A_0^2+6240 (\theta_0-\theta) \cos (2 i_0) A_0^2 \nonumber \\
   & + & 9480
   (\theta_0-\theta) \cos (4 i_0) A_0^2-2160 \theta_0 \cos (2 (i_0-2
   \theta_0)) A_0^2 \nonumber \\
   & + & 2160 \theta \cos (2 (i_0-2 \theta_0)) A_0^2+2304
   \theta_0 \cos (2 (i_0-\theta_0)) A_0^2 \nonumber \\
   & - & 2304 \theta \cos (2
   (i_0-\theta_0)) A_0^2+540 \theta_0 \cos (4 (i_0-\theta_0)) A_0^2-540
   \theta \cos (4 (i_0-\theta_0)) A_0^2 \nonumber \\
   & - & 4608 \theta_0 \cos (2 \theta_0)
   A_0^2+4608 \theta \cos (2 \theta_0) A_0^2+3240 \theta_0 \cos (4
   \theta_0) A_0^2 \nonumber \\
   & - & 3240 \theta \cos (4 \theta_0) A_0^2+2304 \theta_0
   \cos (2 (i_0+\theta_0)) A_0^2-2304 \theta \cos (2 (i_0+\theta_0))
   A_0^2 \nonumber \\
   & + & 540 \theta_0 \cos (4 (i_0+\theta_0)) A_0^2-540 \theta \cos (4
   (i_0+\theta_0)) A_0^2-2160 \theta_0 \cos (2 (i_0+2 \theta_0)) A_0^2 \nonumber \\
   & + & 2160
   \theta \cos (2 (i_0+2 \theta_0)) A_0^2-576 \theta_0 \cos (2
   i_0-\theta_0-\theta) A_0^2 \nonumber \\
   & + & 576 \theta \cos (2 i_0-\theta_0-\theta)
   A_0^2+720 \theta_0 \cos (4 i_0-\theta_0-\theta) A_0^2 \nonumber \\
   & - & 720 \theta \cos
   (4 i_0-\theta_0-\theta) A_0^2-12096 \theta_0 \cos
   (\theta_0-\theta) A_0^2+12096 \theta \cos (\theta_0-\theta)
   A_0^2 \nonumber \\
   & - & 8064 \theta_0 \cos (2 i_0+\theta_0-\theta) A_0^2+8064 \theta
   \cos (2 i_0+\theta_0-\theta) A_0^2 \nonumber \\
   & - & 4320 \theta_0 \cos (4
   i_0+\theta_0-\theta) A_0^2+4320 \theta \cos (4
   i_0+\theta_0-\theta) A_0^2 \nonumber \\
   & - & 672 \theta_0 \cos (3 \theta_0-\theta)
   A_0^2+672 \theta \cos (3 \theta_0-\theta) A_0^2-1344 \theta_0 \cos
   (2 i_0+3 \theta_0-\theta) A_0^2 \nonumber \\
   & + & 1344 \theta \cos (2 i_0+3
   \theta_0-\theta) A_0^2+1680 \theta_0 \cos (4 i_0+3
   \theta_0-\theta) A_0^2 \nonumber \\
   & - & 1680 \theta \cos (4 i_0+3
   \theta_0-\theta) A_0^2-1344 \theta_0 \cos (2 i_0-3
   \theta_0+\theta) A_0^2 \nonumber \\
   & + & 1344 \theta \cos (2 i_0-3
   \theta_0+\theta) A_0^2+1680 \theta_0 \cos (4 i_0-3
   \theta_0+\theta) A_0^2 \nonumber \\
   & - & 1680 \theta \cos (4 i_0-3
   \theta_0+\theta) A_0^2-8064 \theta_0 \cos (2 i_0-\theta_0+\theta)
   A_0^2 \nonumber \\
   & + & 8064 \theta \cos (2 i_0-\theta_0+\theta) A_0^2-4320 \theta_0
   \cos (4 i_0-\theta_0+\theta) A_0^2 \nonumber \\
   & + & 4320 \theta \cos (4
   i_0-\theta_0+\theta) A_0^2-288 \theta_0 \cos (\theta_0+\theta)
   A_0^2+288 \theta \cos (\theta_0+\theta) A_0^2 \nonumber \\
   & - & 576 \theta_0 \cos (2
   i_0+\theta_0+\theta) A_0^2+576 \theta \cos (2 i_0+\theta_0+\theta)
   A_0^2 \nonumber \\
   & + & 720 \theta_0 \cos (4 i_0+\theta_0+\theta) A_0^2-720 \theta \cos
   (4 i_0+\theta_0+\theta) A_0^2 \nonumber \\
   & + & 960 \sin (2 (i_0-2 \theta_0)) A_0^2+3672
   \sin (4 i_0-2 \theta_0) A_0^2+3936 \sin (2 (i_0-\theta_0)) A_0^2 \nonumber \\
   & - & 516 \sin (4
   (i_0-\theta_0)) A_0^2-4752 \sin (2 \theta_0) A_0^2+2040 \sin (4 \theta_0)
   A_0^2 \nonumber \\
   & - & 3936 \sin (2 (i_0+\theta_0)) A_0^2+516 \sin (4 (i_0+\theta_0)) A_0^2-3672
   \sin (2 (2 i_0+\theta_0)) A_0^2 \nonumber \\
   & - & 960 \sin (2 (i_0+2 \theta_0)) A_0^2+48 \sin
   (2 i_0-\theta_0-3 \theta) A_0^2-60 \sin (4 i_0-\theta_0-3 \theta)
   A_0^2 \nonumber \\
   & + & 1008 \sin (\theta_0-3 \theta) A_0^2+672 \sin (2 i_0+\theta_0-3
   \theta) A_0^2+360 \sin (4 i_0+\theta_0-3 \theta) A_0^2 \nonumber \\
   & + & 112 \sin (2 i_0+3
   \theta_0-3 \theta) A_0^2-140 \sin (4 i_0+3 \theta_0-3 \theta)
   A_0^2 \nonumber \\
   & + & 240 \sin (2 (i_0-2 \theta)) A_0^2-1772 \sin (4 i_0-2 \theta)
   A_0^2-504 \sin (4 \theta_0-2 \theta) A_0^2 \nonumber \\
   & - & 756 \sin (4 i_0+4 \theta_0-2
   \theta) A_0^2+882 \sin (6 \theta_0-2 \theta) A_0^2+147 \sin (4 i_0+6
   \theta_0-2 \theta) A_0^2 \nonumber \\
   & - & 3056 \sin (2 (i_0-\theta)) A_0^2+264 \sin (4
   (i_0-\theta)) A_0^2+864 \sin (2 i_0-3 \theta_0-\theta) A_0^2 \nonumber \\
   & - & 144 \sin (4
   i_0-3 \theta_0-\theta) A_0^2-60 \sin (2 (i_0-\theta_0-\theta))
   A_0^2 \nonumber \\
   & - & 1728 \sin (2 i_0-\theta_0-\theta) A_0^2-1296 \sin (4
   i_0-\theta_0-\theta) A_0^2-1896 \sin (\theta_0-\theta) A_0^2 \nonumber \\
   & + & 3540
   \sin (2 (\theta_0-\theta)) A_0^2+56 \sin (3 (\theta_0-\theta))
   A_0^2+840 \sin (2 (i_0+\theta_0-\theta)) A_0^2 \nonumber \\
   & + & 6384 \sin (2
   i_0+\theta_0-\theta) A_0^2+846 \sin (2 (2 i_0+\theta_0-\theta))
   A_0^2 \nonumber \\
   & - & 828 \sin (4 i_0+\theta_0-\theta) A_0^2+1008 \sin (2 (i_0+2
   \theta_0-\theta)) A_0^2+3536 \sin (3 \theta_0-\theta) A_0^2 \nonumber \\
   & - & 588
   \sin (2 (i_0+3 \theta_0-\theta)) A_0^2-1184 \sin (2 i_0+3
   \theta_0-\theta) A_0^2 \nonumber \\
   & + & 1720 \sin (4 i_0+3 \theta_0-\theta) A_0^2-3624
   \sin (5 \theta_0-\theta) A_0^2+2640 \sin (2 i_0+5 \theta_0-\theta)
   A_0^2 \nonumber
\end{eqnarray}
\begin{eqnarray}
   & - & 828 \sin (4 i_0+5 \theta_0-\theta) A_0^2+4168 \sin (2 \theta)
   A_0^2-144 \sin (4 \theta) A_0^2 \nonumber \\
   & + & 3056 \sin (2 (i_0+\theta)) A_0^2-264 \sin
   (4 (i_0+\theta)) A_0^2+1772 \sin (2 (2 i_0+\theta)) A_0^2 \nonumber \\
   & - & 2640 \sin (2
   i_0-5 \theta_0+\theta) A_0^2+828 \sin (4 i_0-5 \theta_0+\theta)
   A_0^2 \nonumber \\
   & + & 588 \sin (2 (i_0-3 \theta_0+\theta)) A_0^2+1184 \sin (2 i_0-3
   \theta_0+\theta) A_0^2 \nonumber \\
   & - & 147 \sin (2 (2 i_0-3 \theta_0+\theta))
   A_0^2-1720 \sin (4 i_0-3 \theta_0+\theta) A_0^2 \nonumber \\
   & - & 1008 \sin (2 (i_0-2
   \theta_0+\theta)) A_0^2+756 \sin (2 (2 i_0-2 \theta_0+\theta))
   A_0^2 \nonumber \\
   & - & 840 \sin (2 (i_0-\theta_0+\theta)) A_0^2-6384 \sin (2
   i_0-\theta_0+\theta) A_0^2 \nonumber \\
   & - & 846 \sin (2 (2 i_0-\theta_0+\theta))
   A_0^2+828 \sin (4 i_0-\theta_0+\theta) A_0^2-1440 \sin
   (\theta_0+\theta) A_0^2 \nonumber \\
   & - & 570 \sin (2 (\theta_0+\theta)) A_0^2+60
   \sin (2 (i_0+\theta_0+\theta)) A_0^2+1728 \sin (2
   i_0+\theta_0+\theta) A_0^2 \nonumber \\
   & + & 225 \sin (2 (2 i_0+\theta_0+\theta))
   A_0^2+1296 \sin (4 i_0+\theta_0+\theta) A_0^2+1440 \sin (3
   \theta_0+\theta) A_0^2 \nonumber \\
   & - & 864 \sin (2 i_0+3 \theta_0+\theta) A_0^2+144
   \sin (4 i_0+3 \theta_0+\theta) A_0^2-240 \sin (2 (i_0+2 \theta))
   A_0^2 \nonumber \\
   & - & 112 \sin (2 i_0-3 \theta_0+3 \theta) A_0^2+140 \sin (4 i_0-3
   \theta_0+3 \theta) A_0^2 \nonumber \\
   & - & 672 \sin (2 i_0-\theta_0+3 \theta) A_0^2-360
   \sin (4 i_0-\theta_0+3 \theta) A_0^2-24 \sin (\theta_0+3 \theta)
   A_0^2 \nonumber \\
   & - & 48 \sin (2 i_0+\theta_0+3 \theta) A_0^2+60 \sin (4 i_0+\theta_0+3
   \theta) A_0^2-225 \sin (4 i_0-2 (\theta_0+\theta)) A_0^2 \nonumber \\
   & + & 3456
   \theta_0 X_0 \cos (2 i_0-3 \theta_0) A_0-3456 \theta X_0 \cos (2 i_0-3
   \theta_0) A_0 \nonumber \\
   & + & 1216 Y_0 \cos (2 i_0-3 \theta_0) A_0-3456 \theta_0 X_0 \cos (2
   i_0-\theta_0) A_0 \nonumber \\
   & + & 3456 \theta X_0 \cos (2 i_0-\theta_0) A_0+4416 Y_0 \cos (2
   i_0-\theta_0) A_0-11520 \theta_0 X_0 \cos (\theta_0) A_0 \nonumber \\
   & + & 11520 \theta X_0
   \cos (\theta_0) A_0-2688 Y_0 \cos (\theta_0) A_0-6912 \theta_0 X_0 \cos (3
   \theta_0) A_0 \nonumber \\
   & + & 6912 \theta X_0 \cos (3 \theta_0) A_0-3456 Y_0 \cos (3
   \theta_0) A_0-3456 \theta_0 X_0 \cos (2 i_0+\theta_0) A_0 \nonumber \\
   & + & 3456 \theta X_0
   \cos (2 i_0+\theta_0) A_0+4416 Y_0 \cos (2 i_0+\theta_0) A_0 \nonumber \\
   & + & 3456 \theta_0 X_0
   \cos (2 i_0+3 \theta_0) A_0-3456 \theta X_0 \cos (2 i_0+3 \theta_0) A_0 \nonumber \\
   & + & 1216
   Y_0 \cos (2 i_0+3 \theta_0) A_0-640 Y_0 \cos (2 i_0-3 \theta) A_0+576 Y_0 \cos (2
   i_0-\theta_0-2 \theta) A_0 \nonumber \\
   & - & 2304 Y_0 \cos (\theta_0-2 \theta) A_0-3456 Y_0
   \cos (2 i_0+\theta_0-2 \theta) A_0 \nonumber \\
   & - & 2688 Y_0 \cos (3 \theta_0-2
   \theta) A_0+1344 Y_0 \cos (2 i_0+3 \theta_0-2 \theta) A_0 \nonumber \\
   & + & 7680 \theta_0
   X_0 \cos (2 i_0-\theta) A_0-7680 \theta X_0 \cos (2 i_0-\theta) A_0+6144 Y_0
   \cos (2 i_0-\theta) A_0 \nonumber \\
   & - & 2304 Y_0 \cos (2 i_0-2 \theta_0-\theta) A_0+7680 Y_0
   \cos (2 \theta_0-\theta) A_0 \nonumber \\
   & - & 8448 Y_0 \cos (2 i_0+2 \theta_0-\theta)
   A_0-2304 Y_0 \cos (4 \theta_0-\theta) A_0 \nonumber \\
   & + & 1152 Y_0 \cos (2 i_0+4
   \theta_0-\theta) A_0+9216 \theta_0 X_0 \cos (\theta) A_0-9216
   \theta X_0 \cos (\theta) A_0 \nonumber \\
   & - & 768 Y_0 \cos (3 \theta) A_0+7680
   \theta_0 X_0 \cos (2 i_0+\theta) A_0-7680 \theta X_0 \cos (2
   i_0+\theta) A_0 \nonumber \\
   & + & 6144 Y_0 \cos (2 i_0+\theta) A_0+1152 Y_0 \cos (2 i_0-4
   \theta_0+\theta) A_0 \nonumber \\
   & - & 8448 Y_0 \cos (2 i_0-2 \theta_0+\theta) A_0+7680 Y_0
   \cos (2 \theta_0+\theta) A_0 \nonumber \\
   & - & 2304 Y_0 \cos (2 i_0+2 \theta_0+\theta)
   A_0+1344 Y_0 \cos (2 i_0-3 \theta_0+2 \theta) A_0 \nonumber \\
   & - & 3456 Y_0 \cos (2
   i_0-\theta_0+2 \theta) A_0-1152 Y_0 \cos (\theta_0+2 \theta) A_0 \nonumber \\
   & + & 576 Y_0
   \cos (2 i_0+\theta_0+2 \theta) A_0-640 Y_0 \cos (2 i_0+3 \theta) A_0+1216
   X_0 \sin (2 i_0-3 \theta_0) A_0 \nonumber \\
   & - & 3456 \theta_0 Y_0 \sin (2 i_0-3 \theta_0)
   A_0+3456 \theta Y_0 \sin (2 i_0-3 \theta_0) A_0 \nonumber \\
   & - & 3648 X_0 \sin (2 i_0-\theta_0)
   A_0+10368 \theta_0 Y_0 \sin (2 i_0-\theta_0) A_0 \nonumber \\
   & - & 10368 \theta Y_0 \sin (2
   i_0-\theta_0) A_0-7296 X_0 \sin (\theta_0) A_0+2304 \theta_0 Y_0 \sin
   (\theta_0) A_0 \nonumber \\
   & - & 2304 \theta Y_0 \sin (\theta_0) A_0+3456 X_0 \sin (3
   \theta_0) A_0-6912 \theta_0 Y_0 \sin (3 \theta_0) A_0 \nonumber \\
   & + & 6912 \theta Y_0
   \sin (3 \theta_0) A_0+3648 X_0 \sin (2 i_0+\theta_0) A_0-10368 \theta_0 Y_0
   \sin (2 i_0+\theta_0) A_0 \nonumber
\end{eqnarray}
\begin{eqnarray}
   & + & 10368 \theta Y_0 \sin (2 i_0+\theta_0) A_0-1216 X_0
   \sin (2 i_0+3 \theta_0) A_0 \nonumber \\
   & + & 3456 \theta_0 Y_0 \sin (2 i_0+3 \theta_0) A_0-3456
   \theta Y_0 \sin (2 i_0+3 \theta_0) A_0-640 X_0 \sin (2 i_0-3 \theta) A_0 \nonumber \\
   & + & 576
   X_0 \sin (2 i_0-\theta_0-2 \theta) A_0-2304 X_0 \sin (\theta_0-2
   \theta) A_0 \nonumber \\
   & - & 3456 X_0 \sin (2 i_0+\theta_0-2 \theta) A_0-2688 X_0 \sin (3
   \theta_0-2 \theta) A_0 \nonumber \\
   & + & 1344 X_0 \sin (2 i_0+3 \theta_0-2 \theta)
   A_0+4608 X_0 \sin (2 i_0-\theta) A_0 \nonumber \\
   & - & 7680 \theta_0 Y_0 \sin (2 i_0-\theta)
   A_0+7680 \theta Y_0 \sin (2 i_0-\theta) A_0 \nonumber \\
   & - & 2304 X_0 \sin (2 i_0-2
   \theta_0-\theta) A_0+10752 X_0 \sin (2 \theta_0-\theta) A_0 \nonumber \\
   & - & 3840 X_0
   \sin (2 i_0+2 \theta_0-\theta) A_0+2304 X_0 \sin (4 \theta_0-\theta)
   A_0 \nonumber \\
   & - & 1152 X_0 \sin (2 i_0+4 \theta_0-\theta) A_0-3072 X_0 \sin (\theta)
   A_0+9216 \theta_0 Y_0 \sin (\theta) A_0 \nonumber \\
   & - & 9216 \theta Y_0 \sin (\theta)
   A_0+768 X_0 \sin (3 \theta) A_0-4608 X_0 \sin (2 i_0+\theta) A_0 \nonumber \\
   & + & 7680
   \theta_0 Y_0 \sin (2 i_0+\theta) A_0-7680 \theta Y_0 \sin (2
   i_0+\theta) A_0 \nonumber \\
   & + & 1152 X_0 \sin (2 i_0-4 \theta_0+\theta) A_0+3840 X_0 \sin (2
   i_0-2 \theta_0+\theta) A_0 \nonumber \\
   & - & 7680 X_0 \sin (2 \theta_0+\theta) A_0+2304 X_0
   \sin (2 i_0+2 \theta_0+\theta) A_0 \nonumber \\
   & - & 1344 X_0 \sin (2 i_0-3 \theta_0+2
   \theta) A_0+3456 X_0 \sin (2 i_0-\theta_0+2 \theta) A_0 \nonumber \\
   & + & 1152 X_0 \sin
   (\theta_0+2 \theta) A_0-576 X_0 \sin (2 i_0+\theta_0+2 \theta) A_0+640
   X_0 \sin (2 i_0+3 \theta) A_0 \nonumber \\
   & + & 6144 \theta_0 X_0^2-6144 \theta X_0^2+6144
   \theta_0 Y_0^2-6144 \theta Y_0^2-6144 X_0 Y_0 \cos (2 \theta_0) \nonumber \\
   & + & 6144 X_0 Y_0
   \cos (2 \theta)+3072 X_0^2 \sin (2 \theta_0)-3072 Y_0^2 \sin (2
   \theta_0)-3072 X_0^2 \sin (2 \theta) \nonumber \\
   & + & 3072 Y_0^2 \sin (2
   \theta)).
\end{eqnarray}


\section{Transformation from osculating to mean elements} \label{sec:mean}

From the approximate analytical solutions obtained in the previous section it is possible to obtain an analytic transformation from osculating elements to mean elements. The idea behind this is the following. Section~\ref{sec:solution} provides the osculating variation of all the orbital elements of the problem. Therefore, if we perform the integral of the osculating variation for a complete nodal period, that is, from ($\theta_0-\pi$) to ($\theta_0+\pi$), and the result is averaged for the range of the integral ($2\pi$), an approximate mean value of the orbital elements is obtained. In this regard, it is important to note that the center of the domain of the integrals has been selected to be the initial point to account for the expected secular variation of the orbital elements under the effects of $J_2$. Particularly, the mean value of variable $\phi$ can be approximated by:
\begin{equation}
    \overline{\phi} = \phi_0 + \sum_{n=1}^{m} \overline{\phi_n} J_2^n,
\end{equation}
where:
\begin{equation}
    \overline{\phi_n} = \displaystyle\frac{1}{2\pi} \int_{\theta_0 - \pi}^{\theta_0 + \pi} \phi_n.
\end{equation}
It is worth mentioning that this derivation is also performed for the time evolution of the system. The reason for that is because this mean value is related with the mean anomaly of the orbit, and thus, required if we want to perform a transformation between the osculating argument of latitude to the mean anomaly of the orbit in study. The next subsections present the results of these transformations and they are organized based on the order of the solution. Note that there is no zero order transformation.

\subsection{First order transformation}

The first order transformation can be obtained using the results from Eq.~\eqref{eq:1stsolution}:
\begin{eqnarray} \label{eq:1stmean}
\overline{A_1} & = & 3 A^2 \cos(2 \theta_0) \sin^2(i); \nonumber \\
\overline{X_{1}} & = & \displaystyle\frac{1}{8} A (-7 \cos(3 \theta_0) \sin^2(i) + 3 \cos(\theta_0) (-4 + 5 \sin^2(i))); \nonumber \\
\overline{Y_{1}} & = & \displaystyle\frac{1}{2} A \sin(\theta_0) (-3 + 7 \sin^2(i) \sin^2(\theta_0)); \nonumber \\
\overline{i_1} & = & - \displaystyle\frac{3}{8} A \cos(2 \theta_0) \sin(2 i); \nonumber \\
\overline{\Omega_1} & = & - \displaystyle\frac{3}{2} A \cos(i) \cos(\theta_0) \sin(\theta_0); \nonumber \\
\overline{t_1} & = & \frac{1}{16} \sqrt[4]{\frac{R^6}{A^3 \mu^2}} (3 A (1-5 \cos (2 i))
   \sin (2 \theta_0)+32 X \sin (\theta_0)-32 Y \cos (\theta_0)).
\end{eqnarray}

\subsection{Second order transformation}

The second order transformation can be derived from the equations in Section~\ref{sec:2ndsolution}:
\begin{eqnarray}
\overline{A_2} & = & \displaystyle\frac{1}{16} A^2 \sin^2(i) (38 A + 64 X \cos^3(\theta_0) + A (-74 \cos(2 i) \nonumber \\
   & - & 6 (7 + 9 \cos(2 i)) \cos(2 \theta_0) + 39 \cos(4 \theta_0) \sin^2(i)) - 64 Y \sin^3(\theta_0)); \nonumber \\
\overline{X_{2}} & = & \displaystyle\frac{1}{2048} A (-156 A \cos(2 i - 5 \theta_0) + 81 A \cos(4 i - 5 \theta_0) - 
   472 A \cos(2 i - 3 \theta_0) \nonumber \\
   & + & 128 A \cos(4 i - 3 \theta_0) + 288 X \cos(2 (i - 2 \theta_0)) + 384 X \cos(2 (i - \theta_0)) \nonumber \\
   & - & 1884 A \cos(2 i - \theta_0) + 135 A \cos(4 i - \theta_0) - 2262 A \cos(\theta_0) - 2304 X \cos(2 \theta_0) \nonumber \\
   & - & 464 A \cos(3 \theta_0) - 576 X \cos(4 \theta_0) + 150 A \cos(5 \theta_0) + 384 X \cos(2 (i + \theta_0)) \nonumber \\
   & - & 1884 A \cos(2 i + \theta_0) + 135 A \cos(4 i + \theta_0) + 288 X \cos(2 (i + 2 \theta_0)) \nonumber \\
   & - & 472 A \cos(2 i + 3 \theta_0) + 128 A \cos(4 i + 3 \theta_0) - 156 A \cos(2 i + 5 \theta_0) \nonumber \\
   & + & 81 A \cos(4 i + 5 \theta_0) - 288 Y \sin(2 (i - 2 \theta_0)) + 1536 Y \sin(2 (i - \theta_0)) \nonumber \\
   & - & 576 Y \sin(4 \theta_0) - 1536 Y \sin(2 (i + \theta_0)) + 288 Y \sin(2 (i + 2 \theta_0))); \nonumber \\
\overline{Y_{2}} & = & \displaystyle\frac{1}{2048} A (-288 Y \cos(2 (i - 2 \theta_0)) + 1536 Y \cos(2 (i - \theta_0)) -  1536 Y \cos(2 \theta_0)  \nonumber \\
    & + & 576 Y \cos(4 \theta_0) + 1536 Y \cos(2 (i + \theta_0)) - 288 Y \cos(2 (i + 2 \theta_0))  \nonumber \\
    & + & 156 A \sin(2 i - 5 \theta_0) - 81 A \sin(4 i - 5 \theta_0) + 808 A \sin(2 i - 3 \theta_0)  \nonumber \\
    & - & 248 A \sin(4 i - 3 \theta_0) - 288 X \sin(2 (i - 2 \theta_0)) - 384 X \sin(2 (i - \theta_0))  \nonumber \\
    & + & 1164 A \sin(2 i - \theta_0) + 1629 A \sin(4 i - \theta_0) - 2478 A \sin(\theta_0)  \nonumber \\
    & - & 768 X \sin(2 \theta_0) - 32 A \sin(3 \theta_0) - 576 X \sin(4 \theta_0) + 150 A \sin(5 \theta_0)  \nonumber \\
    & + & 384 X \sin(2 (i + \theta_0)) - 1164 A \sin(2 i + \theta_0) - 1629 A \sin(4 i + \theta_0)  \nonumber \\
    & + & 288 X \sin(2 (i + 2 \theta_0)) - 808 A \sin(2 i + 3 \theta_0) + 248 A \sin(4 i + 3 \theta_0)  \nonumber \\
    & - & 156 A \sin(2 i + 5 \theta_0) + 81 A \sin(4 i + 5 \theta_0)); \nonumber
\end{eqnarray}
\begin{eqnarray}
\overline{i_2} & = & \displaystyle\frac{1}{256} A \sin(
  2 i) (A (76 \cos(2 i) + 12 (7 + 9 \cos(2 i)) \cos(2 \theta_0) \nonumber \\
  & + & 3 (-7 + \cos(2 i)) \cos(4 \theta_0)) - 8 (5 A + 16 X \cos^3(\theta_0) - 16 Y \sin^3(\theta_0))); \nonumber \\
\overline{\Omega_2} & = & \displaystyle\frac{1}{128} A (32 \cos(i) (-9 Y \cos(\theta_0) + Y \cos(3 \theta_0) + 4 X \sin^3(\theta_0)) \nonumber \\
  & + & 42 A (5 \cos(i) + 3 \cos(3 i)) \sin(2 \theta_0) - 3 A (11 \cos(i) + \cos(3 i)) \sin(4 \theta_0)); \nonumber \\
\overline{t_2} & = & \frac{1}{512} \sqrt[4]{\frac{R^6}{A^3 \mu^2}} (6 \sin (2
   \theta_0) (A^2 (132 \cos (2 i)+193 \cos (4 i))+91 A^2 \nonumber \\
   & - & 64 X^2+64
   Y^2)-3 A^2 (-80 \cos (2 i)+43 \cos (4 i)+85) \sin (4
   \theta_0) \nonumber \\
   & - & 48 A X (59 \cos (2 i)+5) \sin (\theta_0)+16 A X (19 \cos (2
   i)-27) \sin (3\theta_0) \nonumber \\
   & + & 48 A Y (17 \cos (2 i)+31) \cos (\theta_0)+16
   A Y (27-19 \cos (2 i)) \cos (3 \theta_0) \nonumber \\
   & + & 768 X Y \cos (2\theta_0)).
\end{eqnarray}


\section{Secular variation} \label{sec:secular}

In this work, we define secular variation as the variation that an orbital element experiences after one complete orbital period. That is, for a general orbital element $\phi$, the secular variation is defined as:
\begin{equation}
    \Delta\left.\phi\right|_{sec} = \phi(\theta = \theta_0 + 2\pi) - \phi(\theta = \theta_0).
\end{equation}
Studying this variation is important to determine the secular change of the orbital elements, to identify the conditions for a frozen orbit, and to find the perturbed period of the orbit for any initial condition of the problem. In the following subsections the results of this secular variation is presented for the case of orbital elements and the time evolution (which represents the period of the orbit). Once this is done, the following sections deal with the most common applications of these secular variations, namely: Section~\ref{sec:frozen} covers the study of the frozen condition, Section~\ref{sec:rgt} covers the case of repeating ground-track orbits, and Section~\ref{sec:sunsync} covers the case of sun-synchronous orbits. 

\subsection{First order secular variation}

The first order secular variation of the orbital elements can be obtained directly from the results in Eq.~\eqref{eq:1stsolution} by imposing $\theta = \theta_0 + 2_\pi$. This leads to the following expressions:
\begin{eqnarray}
\Delta\left.A_1\right|_{sec} & = & 0; \nonumber \\
\Delta\left.X_1\right|_{sec} & = & 0; \nonumber \\
\Delta\left.Y_1\right|_{sec} & = & 0; \nonumber \\
\Delta\left.i_1\right|_{sec} & = & 0; \nonumber \\
\Delta\left.\Omega_1\right|_{sec} & = & -3 A_0 \pi \cos(i_0).
\end{eqnarray}
Note that this result in the right ascension of the ascending node corresponds to the well know expression for the secular variation of this orbital element:
\begin{equation}
    \Delta\left.\Omega\right|_{sec} = -\frac{3}{2} \frac{R^2}{a^2(1-e^2)^2} \sqrt{\frac{\mu}{a^3}} \cos(i_0),
\end{equation}
being the difference, the fact that one of the expressions is related with a time evolution, and the other is related to the evolution of the argument of latitude of the orbit.

\subsection{Second order secular variation}

In a similar process as with the first order, the second order secular variation is obtained by imposing $\theta = \theta_0 + 2_\pi$ in the equations from Section~\ref{sec:2ndsolution}:
\begin{eqnarray}
\Delta\left.A_2\right|_{sec} & = & 0; \nonumber \\
\Delta\left.X_2\right|_{sec} & = & \displaystyle\frac{3}{128} A_0 \pi (-32 Y_0 (3 + 5 \cos(2 i_0)) + 
   3 A_0 (41 + 52 \cos(2 i_0) \nonumber \\
   & + & 35 \cos(4 i_0)) \sin(\theta_0) + 
   28 A_0 (3 + 5 \cos(2 i_0)) \sin^2(i_0) \sin(3 \theta_0)); \nonumber \\
\Delta\left.Y_2\right|_{sec} & = & - \displaystyle\frac{3}{64} A_0 \pi (3 + 5 \cos(2 i_0)) (-16 X_0 + 
   3 A_0 (3 + 5 \cos(2 i_0)) \cos(\theta_0) \nonumber \\
   & + & 14 A_0 \cos(3 \theta_0) \sin^2(i_0)); \nonumber \\
\Delta\left.i_2\right|_{sec} & = & 0; \nonumber \\
\Delta\left.\Omega_2\right|_{sec} & = & \displaystyle\frac{3}{16} A_0^2 \pi (5 \cos(3 i_0) + \cos(i_0) (7 - 60 \cos(2 \theta_0) \sin^2(i_0))).
\end{eqnarray}
It is interesting to note that the apsidal precession is starting to get represented in the secular variation when applying a second order solution using this perturbation approach (it was still not present in the first order solution). On the other hand, there is an additional term appearing in the secular variation of the right ascension of the ascending node. Since the first order solution was written in terms of osculating elements, this term should also be regarded as part of the transformation from osculating to mean elements to adjust for the long term effects in this secular variation.

\subsection{Orbital period} \label{sec:period}

As done for the case of orbital elements, it is possible to find the secular variation of the time evolution of the system after one complete orbital revolution. In this case, we will use the more common notation for orbital period $T$, which using the power series expansion is represented as:
\begin{equation}
    T \approx T_0 + J_2 T_1 + J_2^2 T_2,
\end{equation}
where the zero order period is:
\begin{equation}
    T_0 = 2 \pi \sqrt[4]{\frac{R^6}{A_0^3 \mu^2}};
\end{equation}
the first order correction is:
\begin{equation} \label{eq:1stperiod}
    T_1 = - \displaystyle\frac{3}{2} A_0 \pi \sqrt[4]{\frac{R^6}{A_0^3 \mu^2}} (2 + 4 \cos(2 i_0) + 3 \cos(2 \theta_0) \sin^2(i_0));
\end{equation}
and the second order correction is:
\begin{eqnarray}
    T_2 & = &  \displaystyle\frac{3}{512} \pi \sqrt[4]{\frac{R^6}{A_0^3 \mu^2}} (-14 A_0^2 + 512 X_0^2 + 512 Y_0^2 - 824 A_0^2 \cos(2 i_0) \nonumber \\
    & + & 70 A_0^2 \cos(4 i_0) + 
    288 A_0 X_0 \cos(2 i_0 - 3 \theta_0) - 180 A_0^2 \cos(2 (i_0 - 2 \theta_0)) \nonumber \\
    & + & 
    200 A_0^2 \cos(4 i_0 - 2 \theta_0) + 32 A_0^2 \cos(2 (i_0 - \theta_0)) + 
    45 A_0^2 \cos(4 (i_0 - \theta_0)) \nonumber \\
    & + & 352 A_0 X_0 \cos(2 i_0 - \theta_0) - 192 A_0 X_0 \cos(\theta_0) - 
    464 A_0^2 \cos(2 \theta_0) \nonumber \\
    & - & 576 A_0 X_0 \cos(3 \theta_0) + 270 A_0^2 \cos(4 \theta_0) + 
    32 A_0^2 \cos(2 (i_0 + \theta_0)) \nonumber \\
    & + & 45 A_0^2 \cos(4 (i_0 + \theta_0)) + 
    352 A_0 X_0 \cos(2 i_0 + \theta_0) + 200 A_0^2 \cos(2 (2 i_0 + \theta_0)) \nonumber \\
    & - & 
    180 A_0^2 \cos(2 (i_0 + 2 \theta_0)) + 288 A_0 X_0 \cos(2 i_0 + 3 \theta_0) \nonumber \\
    & - & 
    288 A_0 Y_0 \sin(2 i_0 - 3 \theta_0) + 224 A_0 Y_0 \sin(2 i_0 - \theta_0) + 960 A_0 Y_0 \sin(\theta_0) \nonumber \\
    & - & 
    576 A_0 Y_0 \sin(3 \theta_0) - 224 A_0 Y_0 \sin(2 i_0 + \theta_0) + 288 A_0 Y_0 \sin(2 i_0 + 3 \theta_0)).
\end{eqnarray}
In here, it is important to note that this solution depends on the position because it also represents the transformation from osculating to mean orbital elements appearing as part of this secular effect. This is most clearly seen in Eq.~\eqref{eq:1stperiod} when compared with the first order transformation of variable $A$ between osculating to mean elements (Eq.~\eqref{eq:1stmean}).


\section{Frozen condition} \label{sec:frozen}

For the purpose of this work, we use the definition of frozen condition where the secular variation of the orbital elements $A$, $i$, $X$ and $Y$ are zero after one complete orbital revolution. In other words, this condition is equivalent to:
\begin{equation}
    \Delta\left.A\right|_{sec} =  \Delta\left.i\right|_{sec} =  \Delta\left.X\right|_{sec} =  \Delta\left.Y\right|_{sec} = 0,
\end{equation}
which applied to the first and second order secular variations obtained in Section~\ref{sec:secular}, leads to the following initial conditions in $X$ and $Y$:
\begin{eqnarray} \label{eq:frozen_circ}
    X_0 & = & \displaystyle\frac{1}{16} A_0 (9 \cos(\theta_0) + 15 \cos(2 i_0) \cos(\theta_0) + 14 \cos(3 \theta_0) \sin^2(i_0)); \nonumber \\
    Y_0 & = & \displaystyle\frac{1}{16} A_0 \sin(\theta_0) (10 + 14 \cos(2 i_0) - 7 \cos(2 (i_0 - \theta_0))\nonumber \\
    & + & 14 \cos(2 \theta_0) - 7 \cos(2 (i_0 + \theta_0))),
\end{eqnarray}
which as can be seen have no singularities. This means that there always exist a unique frozen orbit with small eccentricity for any combination of magnitude of the angular momentum (which is directly related with the orbital parameter $A$), and inclination. 

Additionally, this result can be used in combination with Section~\ref{sec:mean} to obtain the mean values of the components of the eccentricity vector for frozen orbits. Particularly:
\begin{eqnarray}
    \overline{e_x} & = & -\displaystyle\frac{1}{256} J_2^2 A_0^2 \sin (\theta_0) (-156 \cos (2 (i_0-2 \theta_0))-212 \cos (4 i_0-2 \theta_0)\nonumber \\
    & + & 208 \cos (2 (i_0-\theta_0))+39 \cos (4(i_0-\theta_0))+208 \cos (2 (i_0+\theta_0))\nonumber \\
    & + & 39 \cos (4 (i_0+\theta_0))-212\cos (2 (2 i_0+\theta_0))-156 \cos (2 (i_0+2 \theta_0))\nonumber \\
    & + & 280 \cos (2 i_0)+346 \cos (4 i_0)+8 \cos (2 \theta_0)+234 \cos (4 \theta_0)+526); \nonumber \\
    \overline{e_y} & = & -\displaystyle\frac{1}{256} J_2^2 A_0^2 \cos (\theta_0) (-156 \cos (2 (i_0-2 \theta_0))-260 \cos (4 i_0-2 \theta_0)\nonumber \\
    & + & 400 \cos (2 (i_0-\theta_0))+39 \cos (4 (i_0-\theta_0))+400 \cos (2 (i_0+\theta_0))\nonumber \\
    & + & 39 \cos (4 (i_0+\theta_0))-260 \cos (2 (2 i_0+\theta_0))-156 \cos (2 (i_0+2 \theta_0)) \nonumber \\
    & + & 152 \cos (2 i_0)+314 \cos (4 i_0)-280 \cos (2 \theta_0)+234 \cos (4 \theta_0)+686);
\end{eqnarray}
where it is interesting to observe that the mean value of both mean components is in the order of magnitude of $J_2^2$. Note also that these mean values corroborate the fact that these low eccentric frozen orbits are unique for each combination of magnitude of the angular momentum and inclination. 

\begin{figure}[h!]
	\centering
	{\includegraphics[width = 0.8\textwidth]{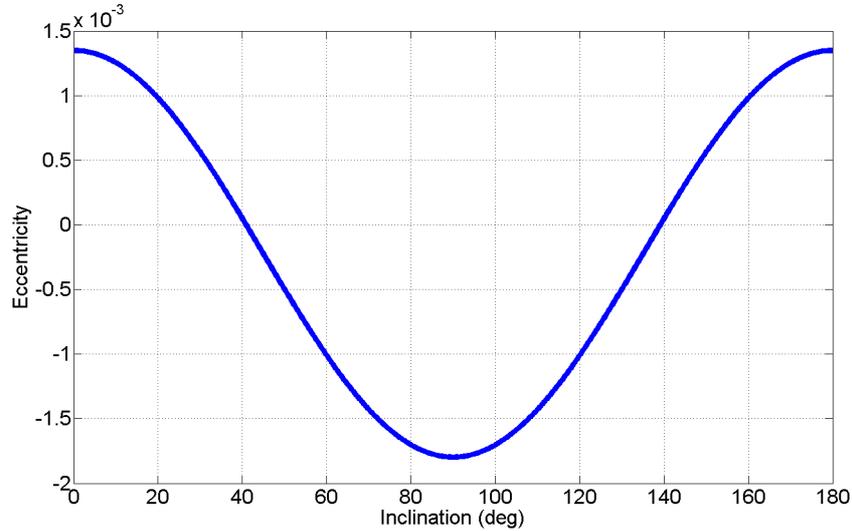}}
	\caption{Relation between initial osculating inclination and eccentricity.}
	\label{fig:frozen_inc_ey}
\end{figure}

Figure~\ref{fig:frozen_inc_ey} shows the relation between the initial osculating values of inclination and $y$ component of the eccentricity vector when considering an orbit with $A_0 = 0.8302$. This represents a set of near circular orbits at 7000 km of radius orbiting about the Earth. As can be seen, there is a unique frozen orbit for each value of inclination, not existing any singularity in their distribution.

\section{Repeating ground-track orbits} \label{sec:rgt}

Since the perturbed nodal period of the orbit is known from Section~\ref{sec:period} as well as the evolution of all the orbital elements, it is possible to define frozen orbits that also maintain their repeating ground-track property under the effect of $J_2$. Particularly, we can define the period of repetition of the ground-track as:
\begin{equation}
    T_c = N_d T_{\omega G} = N_p T,
\end{equation}
where $T_{\omega G}$ is the nodal period of Greenwich, $T$ is the perturbed nodal period of the orbit, and $N_d$ and $N_p$ are the number of rotations of the Earth and the orbiting object in order to repeat the ground-track cycle. Additionally, we know that the  nodal period of Greenwich under orbital perturbations can be expressed as:
\begin{equation}
    T_{\omega G} = \displaystyle\frac{2\pi}{\omega_{\oplus} - \dot{\Omega}},
\end{equation}
where $\omega_{\oplus}$ is the main celestial body's spin rate about its axis, and $\dot{\Omega}$ represents the secular rate of change of the right ascension of the ascending node of the orbit. This means that $\dot{\Omega}$ can be represented as:
\begin{equation}
    \dot{\Omega} \approx \displaystyle\frac{\Delta\left.\Omega\right|_{sec}}{T},
\end{equation}
which introduced in the previous expressions leads to the following relation:
\begin{equation}
    2\pi N_d = N_p (\omega_{\oplus} T - \Delta\left.\Omega\right|_{sec}).
\end{equation}
This is a non-linear differential equation that does not have a general solution. Note that since the frozen condition also depends on the initial conditions, this is in fact a system of non-linear equations (the one from the repeating ground-track condition in addition to the equations related with the frozen condition). However, numerical solutions can be obtained using iterative approaches such as Householder's method, Broyden's method, or any other root-finding algorithm using as a starting point of iteration the unperturbed solution.


\section{Sun-synchronous orbits}  \label{sec:sunsync}

Similarly to the case of repeating ground-track orbits, it is also possible to impose the sun-synchronous condition using the results already obtained in the previous sections. Particularly, a sun-synchronous orbit requires a complete rotation of the right ascension of the ascending node after one sidereal year ($t_y$). In particular:
\begin{equation}
    \displaystyle\frac{2\pi}{t_y} \approx \dot{\Omega} \approx \displaystyle\frac{\Delta\left.\Omega\right|_{sec}}{T},
\end{equation}
in other words:
\begin{equation}
    2\pi T - t_y \Delta\left.\Omega\right|_{sec} = 0,
\end{equation}
which as before is a non-linear expression that can be solved numerically using a root-finding algorithm as mentioned above.

Additionally, the sun-synchronous condition can also be combined with the repeating ground-track condition (as in many Earth observation missions) generating a more complex system of non-linear equations. This system includes, in general, the conditions of repeating ground-track orbits, the sun-synchronous condition and the frozen condition. Nevertheless, this system can be solved numerically by iterative methods. An example of that this kind of approach can be seen in Ref.~\cite{tesis}.


\section{Examples of application}  \label{sec:examples}

In this section, three examples of application are presented to show the performance of the proposed methodologies to study frozen orbits. Particularly, the first example covers a near circular orbit, while the other two examples deal with the eccentric orbits discussed in Section~\ref{sec:critical}. All the examples provided assume a motion about the Earth.

\subsection{Near circular orbit}

\begin{figure}[h!]
	\centering
	{\includegraphics[width = 0.95\textwidth]{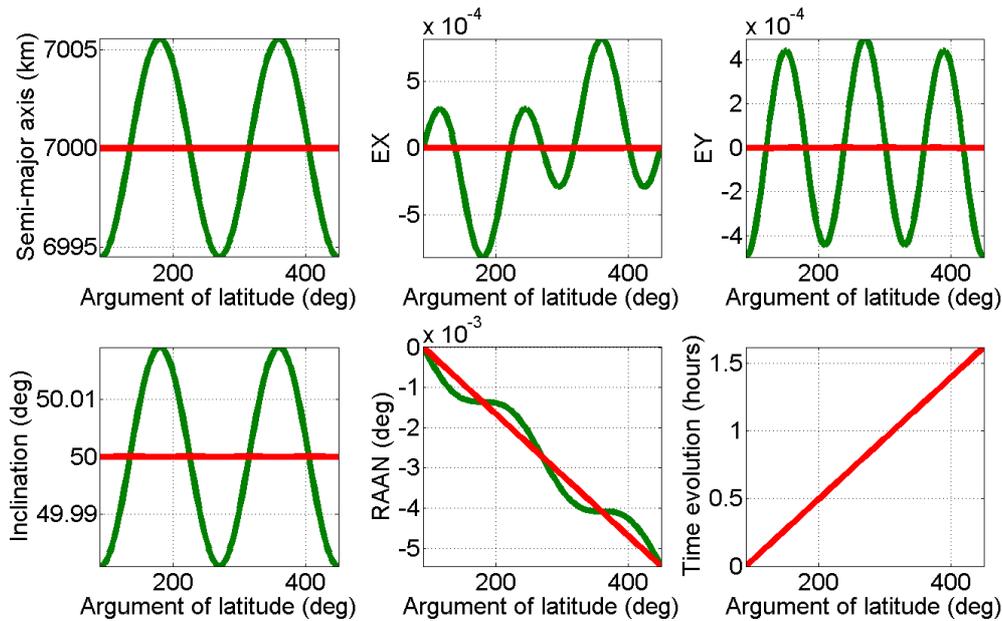}}
	\caption{Osculating and mean evolution of a near circular frozen orbit.}
	\label{fig:mean_50deg}
\end{figure}

For this example, a frozen orbit at mean inclination of 50 deg and mean semi-major axis of 7000 km is selected. Figure~\ref{fig:mean_50deg} shows the evolution of the orbital elements as well as the transformation from osculating to mean elements included in this work. This transformation has been applied on the numerical propagation. As can be seen, the semi-major axis, inclination, and the two components of the eccentricity vector are completely periodic. This can be also seen in the fact that the mean evolution of these variables is an horizontal line, that is, the mean value of these orbital elements do not have any secular component.

\begin{figure}[h!]
	\centering
	{\includegraphics[width = 0.75\textwidth]{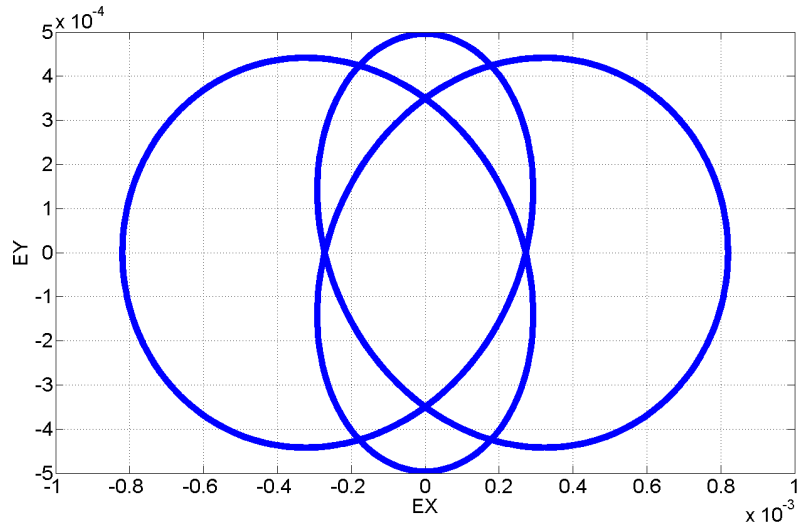}}
	\caption{Osculating value of eccentricity for a near circular frozen orbit.}
	\label{fig:circ_ecc}
\end{figure}

Additionally, Fig.~\ref{fig:circ_ecc} shows a numerical propagation of the eccentricity for a length of one year using the initial conditions provided by Eq.~\ref{eq:frozen_circ}. As can be seen, the orbit maintains its argument of periapsis even after this long-term propagation.

\begin{figure}[h!]
	\centering
	{\includegraphics[width = 0.95\textwidth]{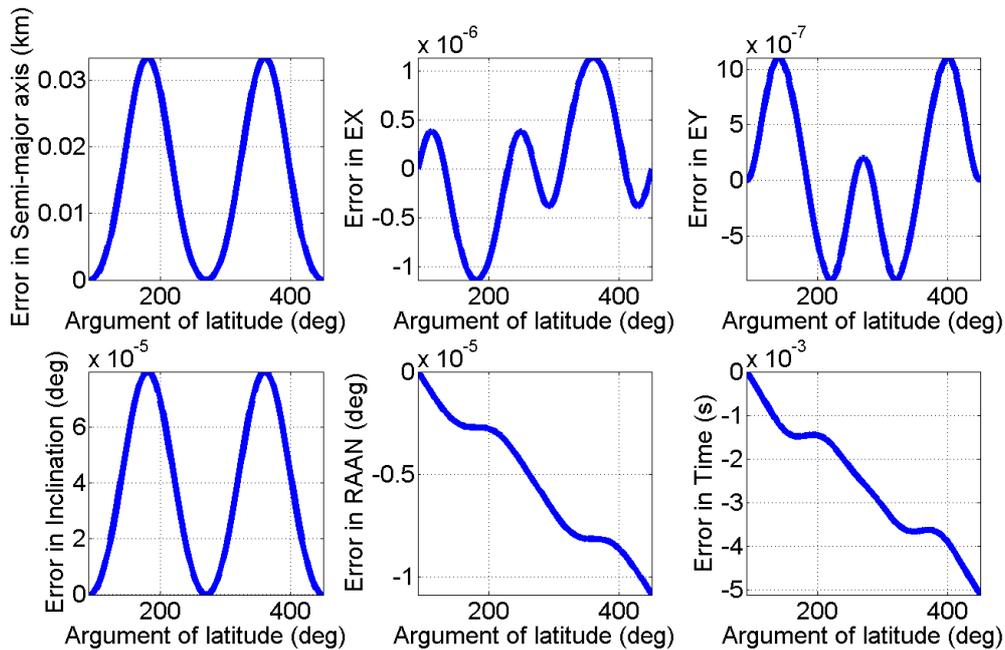}}
	\caption{First order error of a near circular frozen orbit.}
	\label{fig:o1_50deg}
\end{figure}

Another interesting metric to study is the error associated with the approximate analytical solution provided in Section~\ref{sec:solution}. Figure~\ref{fig:o1_50deg} shows the error of the first order solution compared with the numerical solution. This error corresponds to a maximum error in position of less than 63 meters for one orbital period. One of the interesting properties to note in this solution is that the error in semi-major axis, inclination and the two components of the eccentricity gets close to zero after one orbital propagation. This is due to the periodic nature of these orbital elements when dealing with frozen orbits.

\begin{figure}[h!]
	\centering
	{\includegraphics[width = 0.95\textwidth]{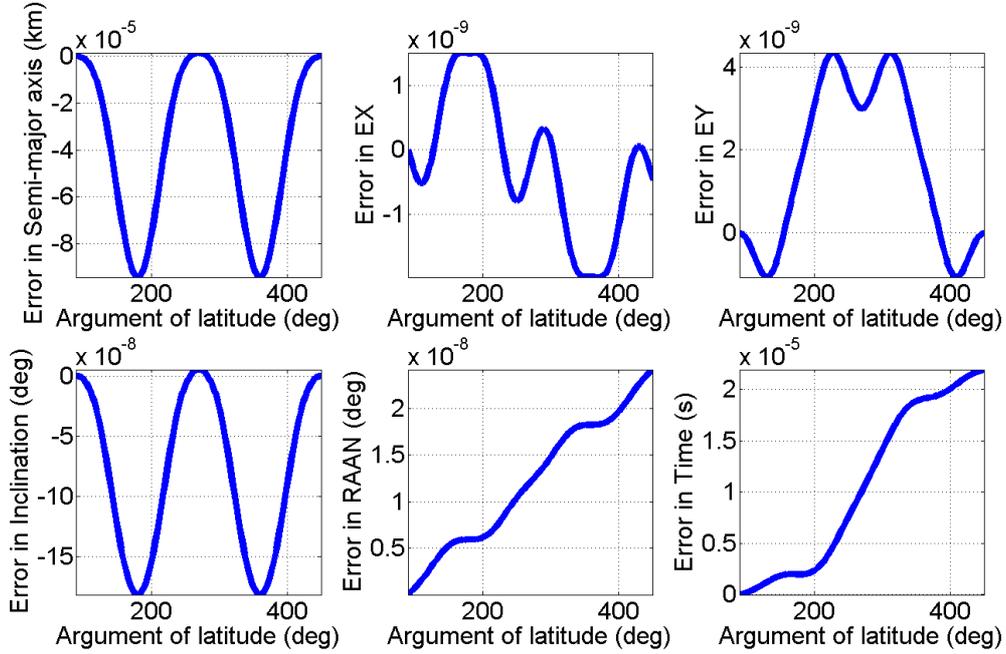}}
	\caption{Second order error of a near circular frozen orbit.}
	\label{fig:o2_50deg}
\end{figure}

\begin{figure}[h!]
	\centering
	{\includegraphics[width = 0.95\textwidth]{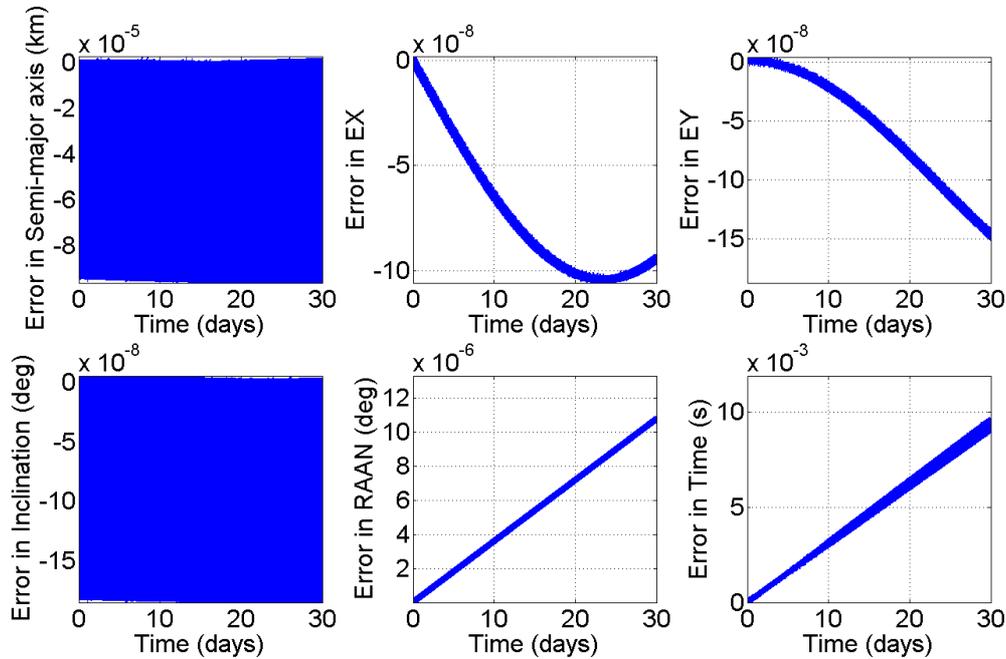}}
	\caption{Second order error in a long-term propagation of a near circular frozen orbit.}
	\label{fig:o2_50deg_long_days}
\end{figure}

This result can be improved using the second order solution. To that end, Fig.~\ref{fig:o2_50deg} shows the error associated with this approximate solution. This error corresponds to a maximum error in position of 15 cm. As can be seen, and compared with the first order solution, all orbital elements improve their error in two orders of magnitude, in fact, close to three orders of magnitude, which is the behaviour that is expected using the perturbation method used. Additionally, the long-term performance of this solution can be also studied. In that regard, Fig.~\ref{fig:o2_50deg_long_days} shows the error associated with this propagation. As expected, the error increases over time to a maximum error in position of 75 meters after 30 days of propagation. 

\begin{figure}[h!]
	\centering
	{\includegraphics[width = 0.95\textwidth]{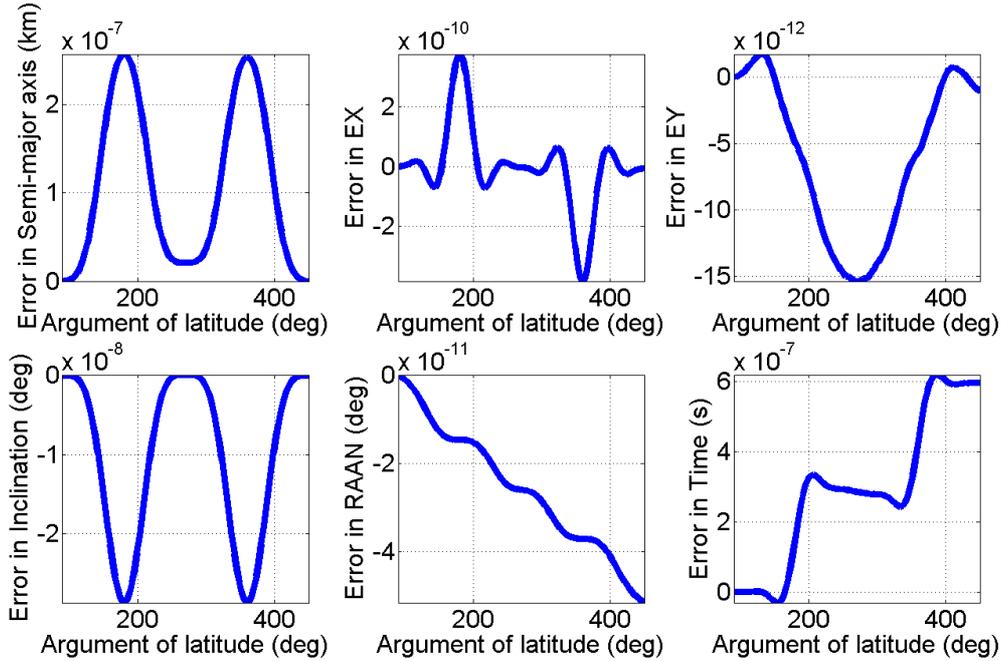}}
	\caption{Third order error of a near circular frozen orbit.}
	\label{fig:o3_50deg}
\end{figure}

Similarly, and due to the simplicity of the perturbation method, it is also possible to generate a third order analytical approximation. The error associated with this solution is provided in Fig.~\ref{fig:o3_50deg}. This corresponds to a maximum error in position of less than 3 mm. As in the previous case, the error has improved again approximately two orders of magnitude with respect to the second order solution, which shows the convergence of the power series to the real solution to the system.

\subsection{Eccentric orbit with $\mathcal{O}(e_{x0}) = \mathcal{O}(J_2)$}

On the other hand, we can study the performance of the solution for eccentric orbits close to the critical inclination. With that goal in mind, in this example an orbit with initial osculating eccentricity of 0.2 and whose periapsis is located at 650 km of altitude over Earth's surface is selected. Particularly, $A_0 = 0.5719$, $i_0 = 63.4235$ deg, $e_{x0} = 0$, $e_{y0} = 0.2$, $\Omega_0 = 0$ deg, and $\theta_0 = 90$ deg. This means that even if the second order solution allows us to define $e_{x0}$ freely as long as it fulfills $\mathcal{O}(e_{x0}) = \mathcal{O}(J_2)$, its value has be chosen such that the orbit is symmetric with respect to the vertical axis of the Earth.

\begin{figure}[h!]
	\centering
	{\includegraphics[width = 0.95\textwidth]{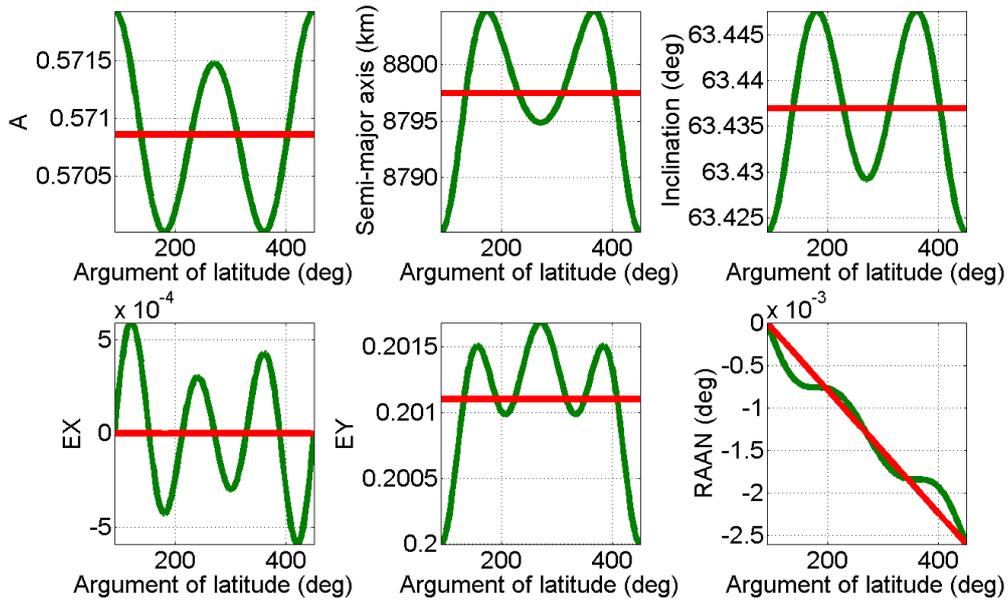}}
	\caption{Osculating and mean evolution of a frozen orbit with $\mathcal{O}(e_{x0}) = \mathcal{O}(J_2)$.}
	\label{fig:mean_eccentric_frozen_90}
\end{figure}

\begin{figure}[h!]
	\centering
	{\includegraphics[width = 0.75\textwidth]{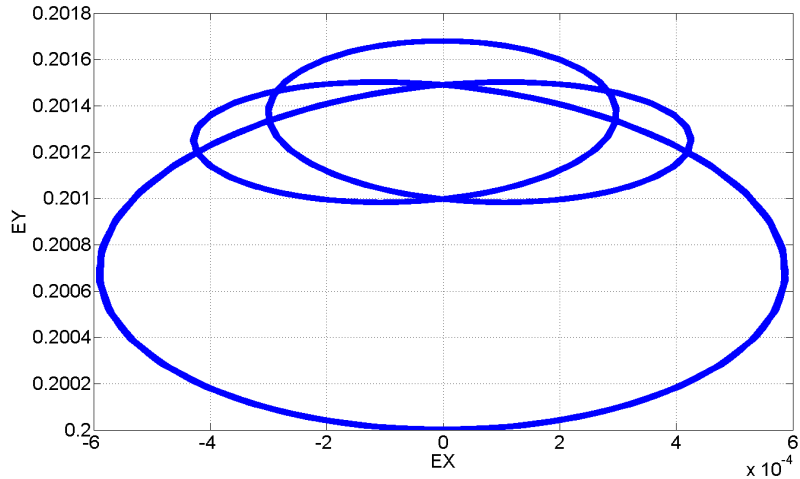}}
	\caption{Osculating eccentricity vector in a one-year propagation of a frozen orbit with $\mathcal{O}(e_{x0}) = \mathcal{O}(J_2)$.}
	\label{fig:eccentric_frozen_90}
\end{figure}

Figure~\ref{fig:mean_eccentric_frozen_90} shows the evolution of the orbital elements as well as their mean value over time. As can be seen, semi-major axis, inclination, and the two components of the eccentricity show a periodic behaviour due to the frozen condition. This can also be seen in the fact that the mean value of these orbital elements remains constant over this propagation. In this regard, it is interesting to study the long term evolution of the eccentricity vector. Figure~\ref{fig:eccentric_frozen_90} shows this evolution using a numerical propagator for a complete one year of propagation. This figure clearly shows that the orbit maintains its frozen condition over the whole propagation time.

\begin{figure}[h!]
	\centering
	{\includegraphics[width = 0.95\textwidth]{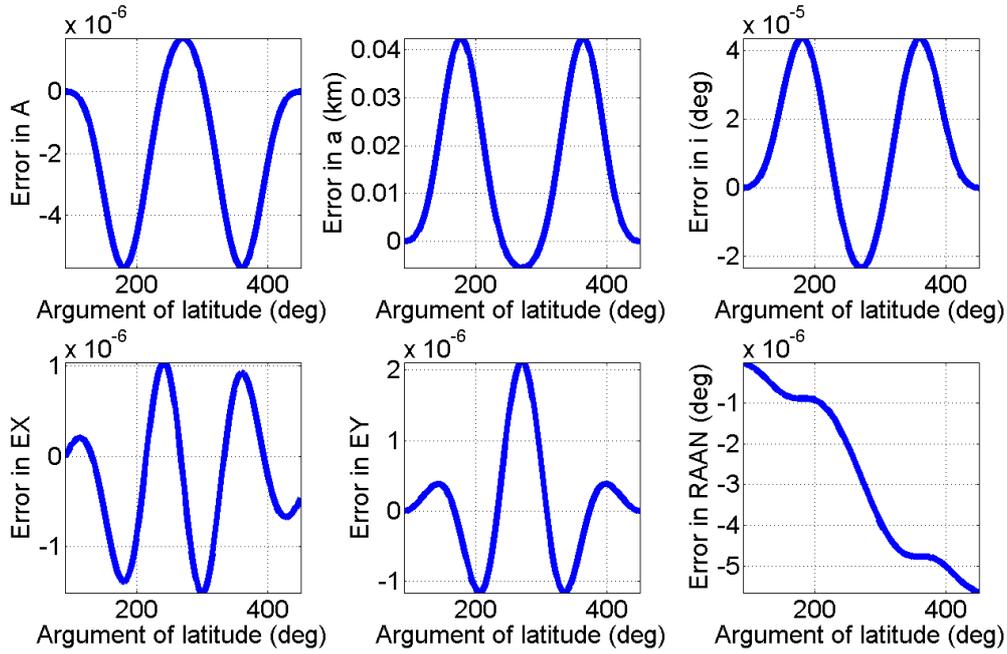}}
	\caption{First order error of a frozen orbit with $\mathcal{O}(e_{x0}) = \mathcal{O}(J_2)$.}
	\label{fig:o1_eccentric_frozen_90}
\end{figure}

Figure~\ref{fig:o1_eccentric_frozen_90} shows the error associated with the first order solution included in Ref.~\cite{meanj2}. As can be seen, even a first order solution provides a very good error accuracy to the system.

\newpage 

\begin{figure}[h!]
	\centering
	{\includegraphics[width = 0.95\textwidth]{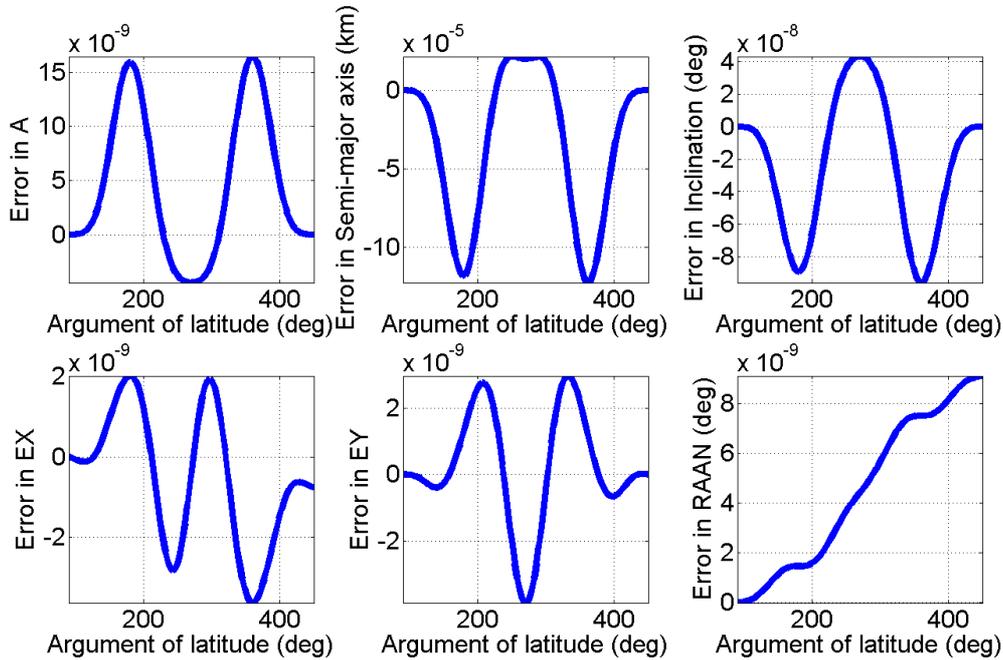}}
	\caption{Second order error of a frozen orbit with $\mathcal{O}(e_{x0}) = \mathcal{O}(J_2)$.}
	\label{fig:o2_eccentric_frozen_90}
\end{figure}

\begin{figure}[h!]
	\centering
	{\includegraphics[width = 0.95\textwidth]{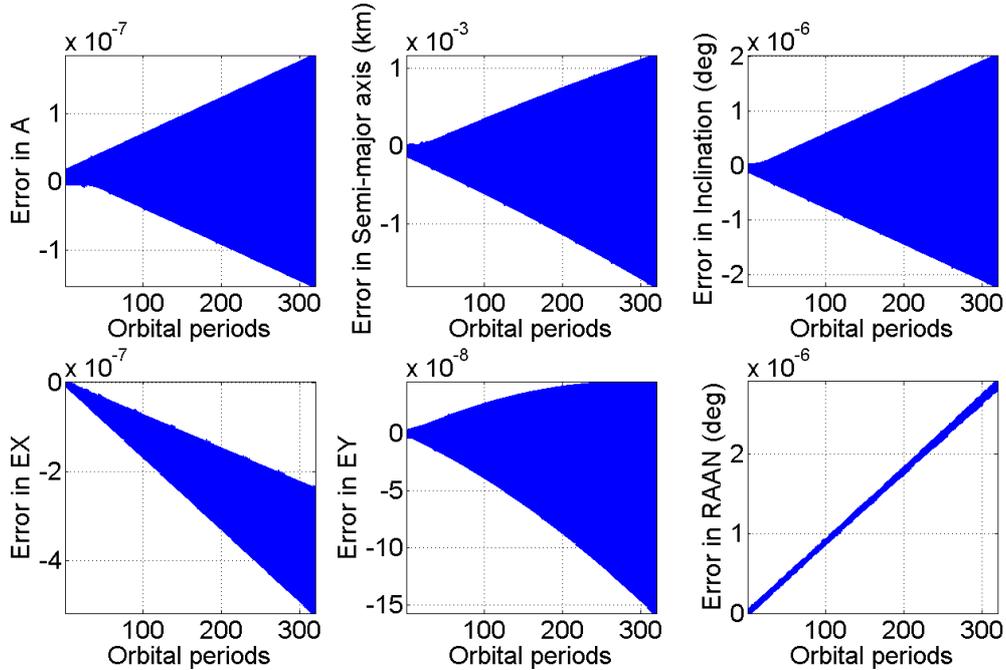}}
	\caption{Second order error in a long-term propagation of a frozen orbit with $\mathcal{O}(e_{x0}) = \mathcal{O}(J_2)$.}
	\label{fig:o2_eccentric_frozen_90_long}
\end{figure}

Similarly, Fig.~\ref{fig:o2_eccentric_frozen_90} shows the error associated with the second order solution. It can be noticed that the error has improved in three orders of magnitude when compared with the first order solution. Additionally, it can be observed that performance is also improved with respect to the examples included in Arnas~\cite{meanj2}. The reason for that is the one that was already included in the mentioned reference, this solution behaves better when closed to the frozen condition due to the lack of control on the perturbed frequency of the solution. This can be seen even clearly in Fig.~\ref{fig:o2_eccentric_frozen_90_long}, where the error associated with a 30-day propagation is presented. These results corroborate that the proposed analytical approximation from Arnas~\cite{meanj2} is especially useful when studying frozen orbits.

\subsection{Eccentric orbit with $\mathcal{O}(e_{y0}) = \mathcal{O}(J_2)$}

As done in the previous example, an orbit with initial osculating eccentricity of 0.2 and whose periapsis is located at 650 km of altitude over Earth's surface is selected, however, this time, the case $\mathcal{O}(e_{y0}) = \mathcal{O}(J_2)$ is considered. Therefore, the following initial conditions are used: $A_0 = 0.5719$, $i_0 = 63.4464$ deg, $e_{x0} = 0.2$, $e_{y0} = J_2$, $\Omega_0 = 0$ deg, and $\theta_0 = 0$ deg. In here, it is important to note that $e_{y0} = J_2$ is chosen to show that the frozen condition also works as expected for non-symmetric orbits with respect to Earth's equator.

\begin{figure}[h!]
	\centering
	{\includegraphics[width = 0.95\textwidth]{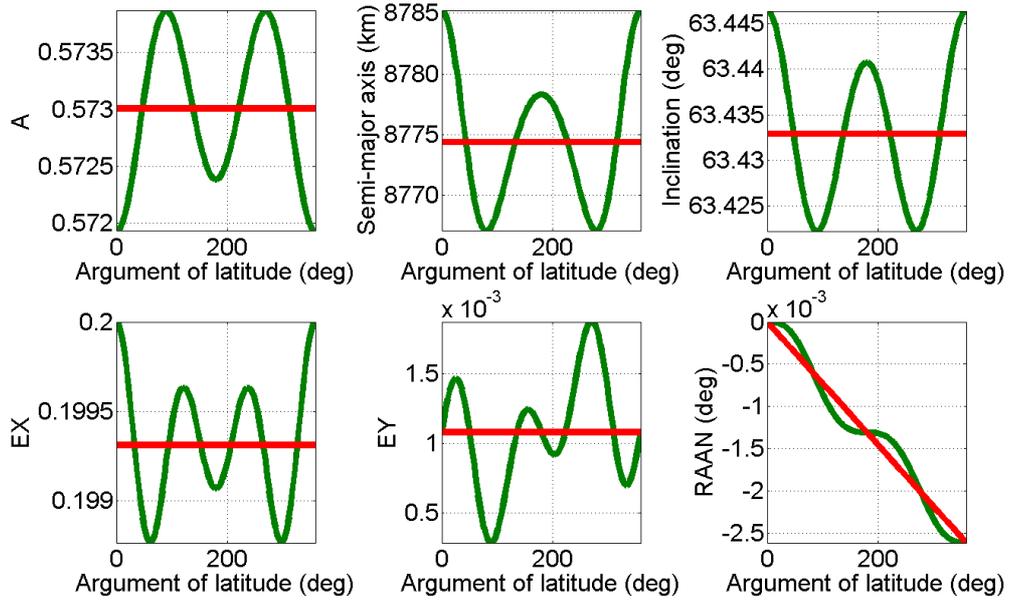}}
	\caption{Osculating and mean evolution of a frozen orbit with $\mathcal{O}(e_{y0}) = \mathcal{O}(J_2)$.}
	\label{fig:mean_eccentric_frozen_00}
\end{figure}

\begin{figure}[h!]
	\centering
	{\includegraphics[width = 0.75\textwidth]{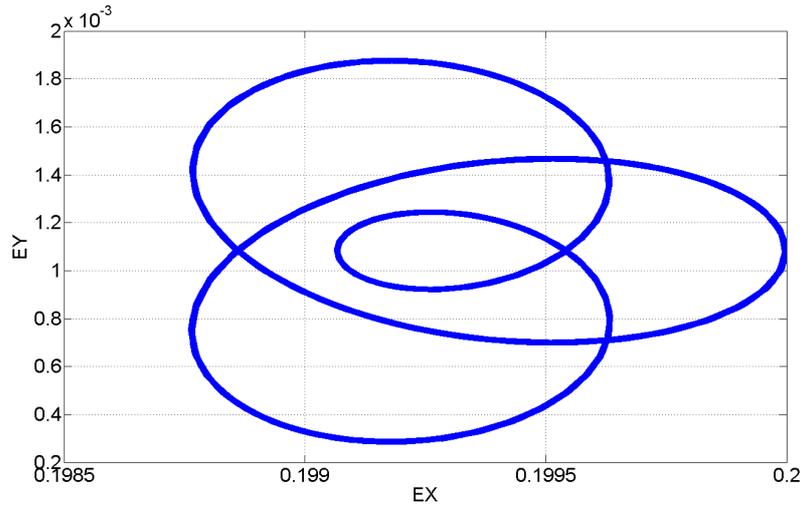}}
	\caption{Osculating eccentricity vector in a one-year propagation of a frozen orbit with $\mathcal{O}(e_{y0}) = \mathcal{O}(J_2)$.}
	\label{fig:eccentric_frozen_00}
\end{figure}

Figure~\ref{fig:mean_eccentric_frozen_00} presents the osculating solution as well as the mean value of the orbital elements. As in the previous examples, the periodic behaviour of $A$, $i$, $e_x$, and $e_y$ is present on the solution. Additionally, Fig.~\ref{fig:eccentric_frozen_00} shows the evolution of the eccentricity vector using a numerical scheme for a propagation lasting one year. This shows that even with the non-symmetrical configuration introduced in the eccentricity (with $e_{y0} = J_2$), the orbit is still behaving as a frozen orbit.

\newpage 

\begin{figure}[h!]
	\centering
	{\includegraphics[width = 0.95\textwidth]{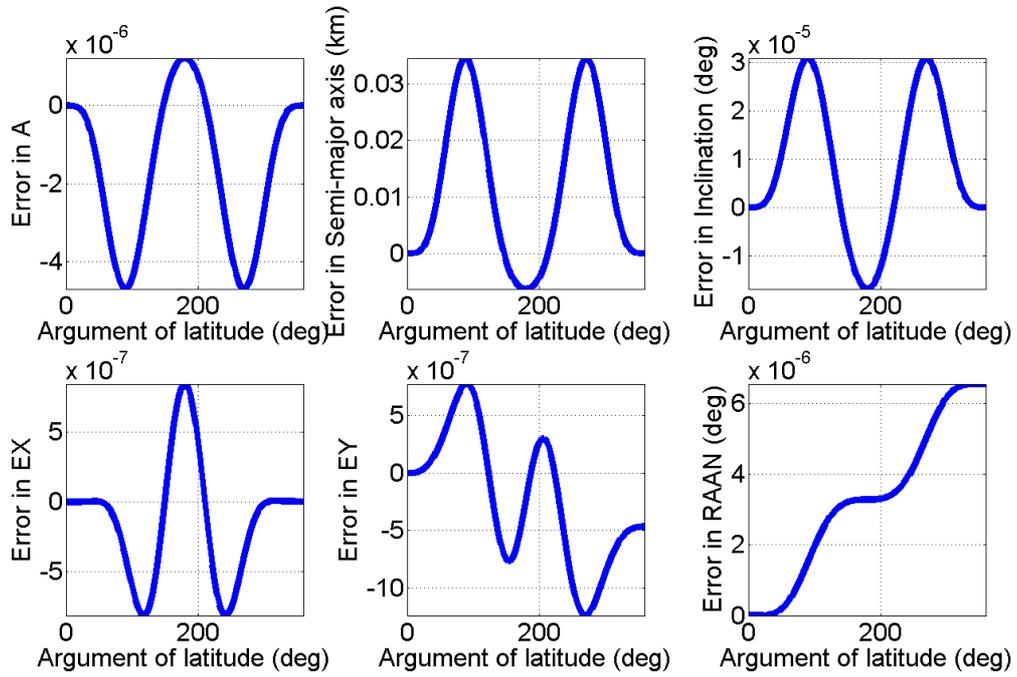}}
	\caption{First order error of a frozen orbit with $\mathcal{O}(e_{y0}) = \mathcal{O}(J_2)$.}
	\label{fig:o1_eccentric_frozen_00}
\end{figure}

\begin{figure}[h!]
	\centering
	{\includegraphics[width = 0.95\textwidth]{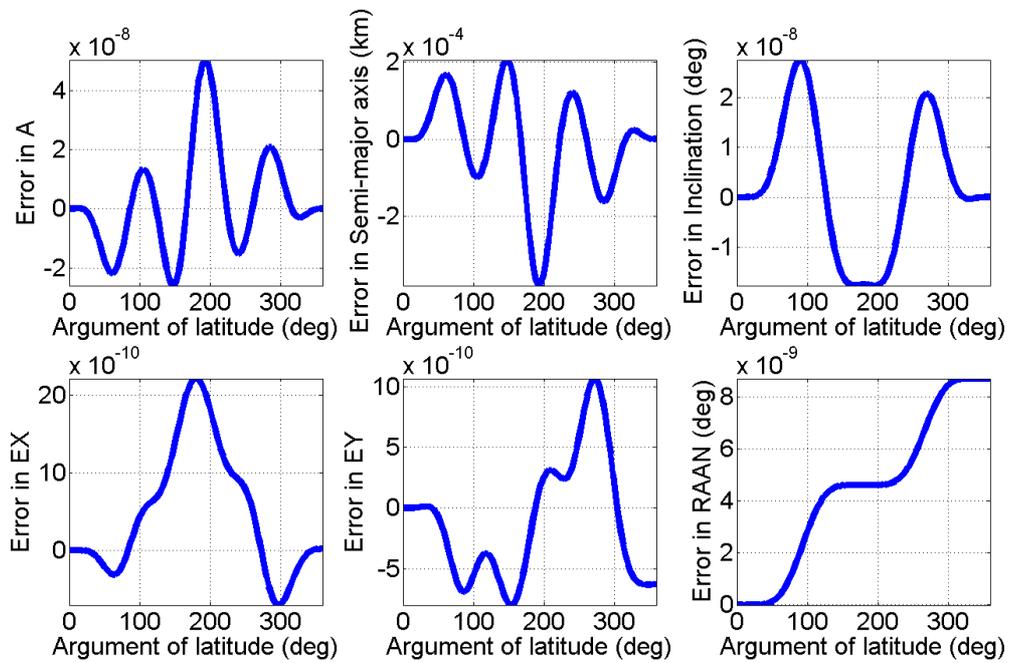}}
	\caption{Second order error of a frozen orbit with $\mathcal{O}(e_{y0}) = \mathcal{O}(J_2)$.}
	\label{fig:o2_eccentric_frozen_00}
\end{figure}

Regarding the error of the methodology, Fig.~\ref{fig:o1_eccentric_frozen_00} shows the one related with the first order solution, while Fig.~\ref{fig:o2_eccentric_frozen_00} does the same for the second order solution. As can be seen, both errors are similar to the ones obtained in Figs.~\ref{fig:o1_eccentric_frozen_90} and~\ref{fig:o2_eccentric_frozen_90} respectively.

Finally, Fig.~\ref{fig:o2_eccentric_frozen_90} shows the long-term error associated with this second order solution. As expected, the performance of this error is comparable to the previous example.

\clearpage

\begin{figure}[h!]
	\centering
	{\includegraphics[width = 0.95\textwidth]{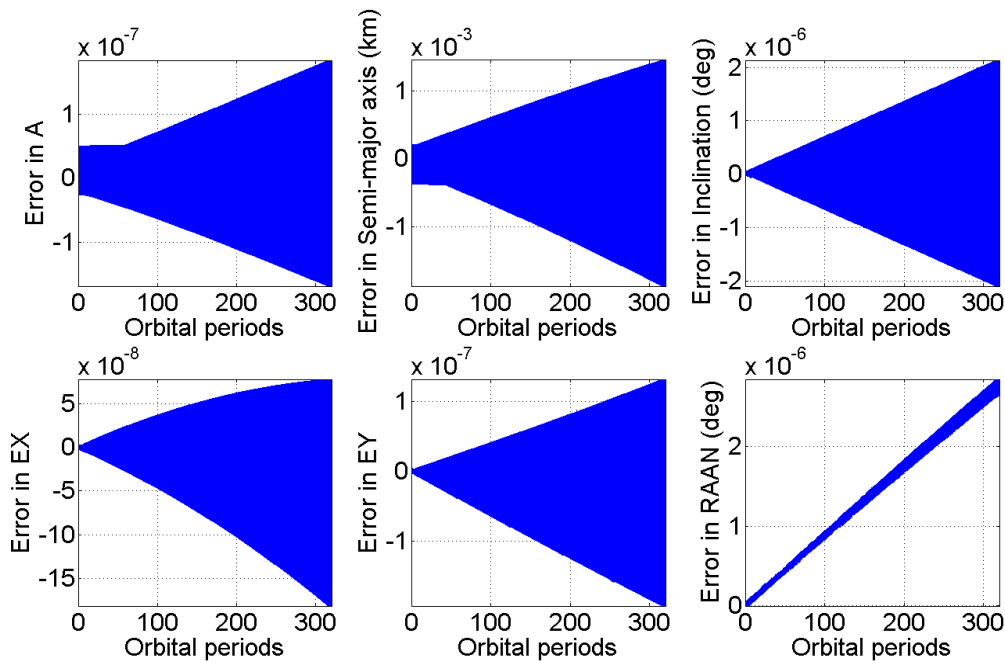}}
	\caption{Second order error in a long-term propagation of a frozen orbit with $\mathcal{O}(e_{y0}) = \mathcal{O}(J_2)$.}
	\label{fig:o2_eccentric_frozen_00_long}
\end{figure}


\section{Conclusions}

This work presents an analytical methodology to study frozen orbits under the effects of $J_2$ perturbation. This is done by generating an approximate analytical solution based on a power series expansion on the osculating orbital elements of the system. This allows to not only find the approximate solution, but also to define a transformation from osculating to mean elements as well as to determine the secular variation of the orbital elements. Two different formulations are used in this paper. The first one does not assume any value on the eccentricity. This allows to study the frozen orbits that appear close to the critical inclination. On the other hand, the second formulation assumes that the magnitude of the eccentricity has the same order of magnitude than $J_2$, therefore, allowing to treat the components of the eccentricity vector as an additional small parameter that can also be expanded in the power series. This alternative formulation allows to study the near circular frozen orbits that appear under $J_2$. Additionally, this second approach drastically reduces the length of the analytical solution when compared with the general solution from the first formulation. This eases the analysis of the resultant expressions and enables the study of the time evolution of the system.    

The results from these perturbation approaches are used to analytically define and study the dynamics of frozen orbits. This includes the family of frozen orbits with magnitude of eccentricity of order $J_2$, as well as the two families of frozen orbits appearing close to the critical inclination. This allows, for instance, to derive a close form expression for the initial orbital elements that provide the frozen condition. In this regard, it is worth mentioning that the second order solutions predicts that when dealing with frozen orbits close to the critical inclination, there is always one component of the eccentricity vector that can be freely set while maintaining the long-term frozen behaviour of the orbit. This result is corroborated by the numerical tests performed. Additionally, the comparison of the approximate analytical solutions with a numerical propagation show that the proposed solutions are able to define accurate frozen orbits and predict their evolution even for very long-term propagations.


\section*{Acknowledgments}

The author wants to thank his parents Mar\'ia Pilar Mart\'inez Ballesteros and  Silvestre Jos\'e Arnas Escart\'in for their support while developing this work. The research contained in this document would not have been possible without their continuous encouragement. Gracias por todo.


\end{document}